\documentclass[reprint,prd,twocolumn]{revtex4}
\usepackage[utf8]{inputenc}
\usepackage[T1]{fontenc}
\usepackage[brazil]{babel}
\usepackage{graphicx}
\usepackage{xcolor}
\usepackage{amsmath}
\usepackage[abbreviations]{siunitx}
\begin{document}
\title{Uma introdução às estrelas estranhas \\
\small{-- An introduction to strange stars -- }}
\author{Victor Paulo Gonçalves}
\email{barros@ufpel.edu.br}
\author{Lucas Lazzari}
\email{lucas.lazzari@outlook.com}
\affiliation{Universidade Federal de Pelotas, Instituto de Física e Matemática, Pelotas, Brasil}
\date{\today}

\begin{abstract}
  A descrição das estrelas de nêutrons mais densas encontradas na Natureza depende da compreensão dos conceitos físicos presentes na Relatividade Geral e na teoria das interações fortes -- a Cromodinâmica Quântica.
  Neste trabalho apresentamos uma revisão dos conceitos básicos necessários para a descrição de estrelas estranhas, formadas pelos quarks {\it up, down} e {\it strange}, a qual é uma das alternativas para a composição de estrelas de nêutrons ultradensas. Iremos revisar a hipótese de Bodmer-Witten-Terazawa que propõe que a matéria estranha é absolutamente estável com relação à matéria nuclear ordinária e discutiremos as propriedades básicas que caracterizam as estrelas estranhas. Nosso objetivo é apresentar os conceitos necessários à compreensão deste tema relevante e atual da área de Astropartículas. \\
  \textbf{Palavras-chave:} Cromodinâmica Quântica; Estrelas estranhas; Estrelas de nêutrons. \\
  
  The description of the heaviest neutron stars 	observed in Nature depends on the understanding of the physical concepts present in General Relativity and  Quantum Chromodynamics. In this work, we review the basic concepts need to describe strange stars, constituted by up, down and strange quarks, which is one of the possible alternatives to describe the general properties of the most dense neutron stars. We will review the Bodmer-Witten-Terazawa hypothesis, which states that strange matter is absolutely stable in relation to ordinary nuclear matter, and discuss the basic properties that characterize strange stars. Our goal is to present the necessary concepts to understand this important theme of Astroparticles.
   \\
  \textbf{Keywords:} Quantum Chromodynamics; Strange stars; Neutron stars.
\end{abstract}
\maketitle

\section{Introdução}
\begin{figure*}[!t]
  \includegraphics[width=\textwidth]{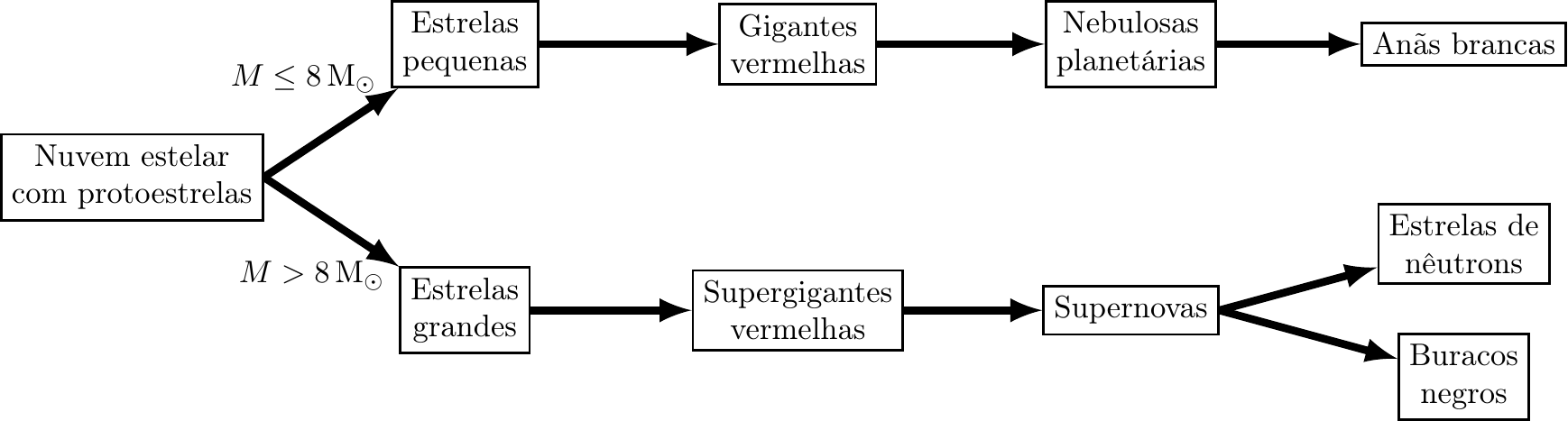}
  \caption{Ilustração esquemática da evolução estelar. Estrelas se formam a partir de protoestrelas geradas pela contração de uma nuvem interestelar de poeira e gás. Estrelas com massas menores do que oito vezes a massa solar (M$_\odot$) tem seu fim em uma anã branca. Estrelas com massas superiores a \num{8}M$_\odot$ explodem em uma supernova, podendo remanescer uma estrela de nêutrons ou um buraco negro.}
  \label{fig:stellar-ev}
\end{figure*}

As estrelas despertaram o interesse da humanidade antes mesmo do advento da ciência. Hoje em dia, sabemos que as estrelas formam-se em uma nuvem de poeira e gás e no seu final, transformam-se em um objeto compacto~\cite{maciel1999}. Através do processo de fusão nuclear, uma estrela passa 90\% de sua vida fundindo hidrogênio em hélio no seu núcleo, na chamada sequência principal~\cite{bandecchi2019}. Esse processo é responsável pela produção de uma pressão expansiva no seu interior, que equilibra a contração gravitacional causada pela própria massa da estrela. Quando o hidrogênio acaba no núcleo, a estrela começa a fundir elementos mais pesados. A partir deste ponto, o futuro da estrela é determinado por sua massa original, tornando-se, no final, um buraco negro ou uma estrela compacta, na forma de uma anã branca ou uma estrela de nêutrons~\cite{maciel1999}, como mostra a Fig.~\ref{fig:stellar-ev}, os quais são os objetos mais densos encontrados na Natureza.

O objetivo deste trabalho é apresentar os conceitos básicos necessários à descrição de estrelas compactas, com particular ênfase na possibilidade de que sejam formadas por quarks. Tal tema, em geral, não é abordado nas disciplinas introdutórias de Astronomia e Astrofísica por envolver conceitos associados à teoria das interações fortes. Neste artigo, buscamos apresentar os conceitos envolvidos na descrição de uma estrela de quarks de uma forma didática, assim como um exemplo simples para uma estrela estranha -- constituída pelos quarks \textit{up}, \textit{down} e \textit{strange}, de forma tal que permita aos estudantes dos semestres iniciais de graduação em Física compreendam este tema tão relevante e atual.

A existência de uma estrela compacta deve-se a Mecânica Quântica. Tem-se que a pressão que sustenta uma estrela compacta é uma consequência do princípio de exclusão de Pauli, onde férmions, como elétrons e nêutrons, não podem ocupar o mesmo estado quântico simultaneamente~\cite{sakurai2013}. Isto leva a formação de um sistema composto por estados de energia degenerados, os quais dão origem a uma pressão de degenerescência. Devido à teoria da relatividade restrita, existe um nível de energia limite, já que nenhuma partícula pode atingir uma velocidade maior do que a da luz. Sendo assim, o limite de energia resulta em uma pressão de degenerescência máxima, para um determinado tipo de partícula. Isso implica na chamada massa de Chandrasekhar~\cite{chandra1931,pinochet2016}, uma massa limite para que elétrons consigam suportar a contração gravitacional, a partir da qual outras partículas, como nêutrons, devem produzir a pressão de degenerescência.

A descrição da evolução e das propriedades das estrelas se dá através das leis da gravitação, isto é, pelas equações do campo gravitacional de Einstein presentes na teoria da relatividade geral~\cite{cheng2005}. Nesta teoria, o espaço-tempo é curvado pela presença de matéria e energia, resultando na ação do campo gravitacional. Entretanto, a Lei da Gravitação Universal proposta por Newton~\cite{newton1999} é aplicável na descrição da maior parte dos objetos astrofísicos como planetas, cometas e asteroides, que apresentam baixas densidades e cujo movimento pode ser corretamente calculado por essa teoria. Assim, a gravitação newtoniana é uma excelente aproximação, que fornece resultados precisos quando os campos gravitacionais são de baixa intensidade~\cite{glendenning2007}. Um dos grandes sucessos da teoria da relatividade geral é a previsão da existência de buracos negros, além de descrever corretamente as propriedades das estrelas mais densas, como é o caso das estrelas de nêutrons. Tal descrição é feita através da equação de Tolman-Oppenheimer-Volkoff (TOV)~\cite{oppenheimer1939,tolman1939}, que é a solução das equações de Einstein do campo gravitacional para uma estrela com simetria esférica, estática e composta por um fluido ideal isotrópico. Essa equação, por sua vez, relaciona a variação da pressão no interior da estrela com a variação do seu raio, podendo ser aplicada para diferentes equações de estado~\cite{glendenning2007}. 

Dentre todos as estrelas que compõem o Universo, as estrelas de nêutrons são as mais densas já observadas. A descrição da sua constituição está diretamente associada à compreensão da teoria das interações fortes, a Cromodinâmica Quântica (QCD, do inglês \textit{Quantum Chromodynamics}), um dos ramos da teoria do Modelo Padrão da Física de Partículas~\cite{moreira2009}. De acordo com esta teoria, nêutrons não são partículas elementares, uma vez que são constituídos por quarks, estes sim ditos elementares. Quarks e antiquarks são definidos a partir da carga elétrica, da massa e do spin, como também da carga de cor, responsável pela interação forte, que é mediada pelos glúons~\cite{silva2008}. As possíveis cargas de cor são o vermelho, o azul e o verde {por convenção}. Além dos quarks e glúons, nenhuma outra partícula possui carga de cor. Existem seis sabores de quarks: \textit{down} (\textit{d}), \textit{up} (\textit{u}), \textit{strange} (\textit{s}), \textit{charm} (\textit{c}), \textit{bottom} (\textit{b}) e \textit{top} (\textit{t}), listados por ordem crescente de massa. Assim como elétrons e nêutrons, quarks são férmions, partículas de spin semi-inteiro sujeitas ao princípio de exclusão de Pauli~\cite{moreira2009}.

As partículas formadas pelos quarks são chamadas de hádrons, compostas de tal forma que sua carga de cor é nula. Consequentemente, estados com três quarks, chamados bárions, são possíveis, pois cada quark possui uma cor resultando em uma combinação ``incolor''. A matéria, como a conhecemos, é dita bariônica, já que prótons e nêutrons são bárions compostos por quarks \textit{u} e \textit{d}. Outros exemplos de bárions são os híperons, que além dos quarks \textit{u} e \textit{d}, possuem também o quark $s$ em diferentes combinações, como \textit{sss} e \textit{uds}. Assim, quarks ganham outra propriedade, o número bariônico igual a um terço, já que os bárions possuem número bariônico igual a um. Devido ao comportamento da interação forte (que cresce com a distância) e de quarks nunca terem sido detectados livres, {postula-se} o conceito de confinamento dos quarks no interior dos hádrons. Por outro lado, a interação entre quarks nas curtas distâncias associadas ao interior dos hádrons é desprezível, levando ao conceito de liberdade assintótica, ou seja, que quarks se comportam como partículas livres no interior dos hádrons~\cite{silva2008}.

Na QCD, a intensidade da interação depende das distâncias entre as partículas. Para distâncias suficientemente pequenas a intensidade da interação será baixa e um tratamento perturbativo é válido, caso contrário, a teoria de perturbação não pode ser aplicada. A caracterização dos hádrons em grandes escalas espaciais, como aquelas presentes em estrelas, é de caráter não perturbativo, com soluções difíceis de serem obtidas. Portanto, modelos fenomenológicos, inspirados pelos resultados da QCD na rede~\cite{wong1994}, são úteis na descrição dos hádrons e também para tratar estados exóticos da matéria.  
{Durante as últimas décadas, diversos modelos foram propostos considerando diferentes tratamentos e aproximações para  as propriedades básicas das interações fortes. Neste trabalho, para fins ilustrativos, iremos considerar o modelo fenomenológico  proposto no \textit{Massachusetts Institute of Technology} (MIT), que é conhecido como modelo de sacola do MIT~\cite{chodos1974,fune2012}. Embora ultrapassado, este modelo ainda é 
 amplamente utilizado na literatura,  pois reproduz de uma forma simples os conceitos de confinamento e liberdade assintótica. Neste modelo, os hádrons são pensados como uma sacola na qual estão contidos os quarks. O movimento dos quarks livres no interior da sacola gera uma pressão, que é contrabalanceada por uma pressão do vácuo, chamada pressão de sacola.}

Dentro do modelo de sacola do MIT, entende-se que existem possíveis configurações nas quais a pressão exercida pelos quarks ultrapassa a pressão de sacola, levando ao rompimento desta e produzindo uma matéria de quarks livres, chamada plasma de quarks e glúons (QGP, do inglês \textit{Quark-Gluon Plasma})~\cite{wong1994}. Isto pode ocorrer em situações exóticas de altas temperaturas e/ou densidades bariônicas elevadas~\cite{nyiri2001}. 
Entretanto, de acordo com a hipótese de Bodmer-Witten-Terazawa, inicialmente proposta por Bodmer~\cite{bodmer1971}, reforçada por Witten~\cite{witten1984} e Terazawa~\cite{terazawa1989}, a matéria de quarks livres contendo os quarks \textit{u}, \textit{d} e \textit{s}, chamada de matéria estranha de quarks (SQM, do inglês \textit{Strange Quark Matter}) é o estado fundamental da matéria que interage fortemente, sendo a configuração mais estável existente. Essa hipótese não contradiz o fato da matéria nuclear ser extremamente mais comum no Universo, pois esta seria um estado metaestável com um tempo de vida da ordem de \num{e14} vezes a idade estimada do Universo~\cite{bodmer1971}. {A validade da hipótese de Bodmer-Witten-Terazawa quando se considera modelos mais sofisticados para tratar as propriedades básicas da QCD ainda é um tema de intenso debate \citep{Klaehn:2017mux}.}

Um dos objetivos da QCD é descrever a interação hadrônica nos regimes de alta energia e densidade elevada, como aqueles presentes em uma estrela de nêutrons. Tal descrição permanece um tema de intenso debate. Ainda assim, entende-se que em decorrência das interações fortes, em altas densidades hadrônicas, híperons sejam produzidos~\cite{weber1999}, permitindo que a estrela possua uma camada composta por nêutrons e um núcleo formado por híperons e, até mesmo, de fases mistas. Como existem vários tipos de híperons, diversas combinações de matéria hadrônica são possíveis. Considerando que a densidade presente nas estrelas de nêutrons seja suficiente para que ocorra {uma transição de fase da matéria hadrônica para uma matéria de quarks, espera-se que a estrela no regime de altas densidades bariônicas seja formada por uma matéria de quarks, juntamente a nêutrons e híperons}. Estas seriam as chamadas estrelas híbridas, com a crosta formada por matéria hadrônica e o núcleo formado por quarks ~\cite{weber1999}. As possibilidades mais extremas são a de uma estrela composta inteiramente por uma matéria de quarks, constituída apenas por quarks \textit{u} e \textit{d} ou pela SQM~\cite{weber1999}. Esta última é chamada estrela estranha.

Neste trabalho, focaremos em estrelas estranhas. 
{A descrição destas estrelas e suas propriedades como, por exemplo, a sua massa e seu raio, dependem fortemente da equação de estado considerada para descrever a interação entre os quarks. Como nosso objetivo é ilustrar os conceitos básicos presentes e o procedimento utilizado para determinar a relação massa-raio destas estrelas, iremos assumir que a matéria estranha pode ser representada por um gás de férmions livres relativísticos. A equação de estado será obtida na aproximação para uma matéria de quarks fria, ou seja, à temperatura zero, a partir do modelo de sacola do MIT~\cite{weber1999}. Dentro deste contexto, 
 começaremos por demonstrar que, em altas densidades bariônicas, a SQM é mais estável do que a matéria ordinária e do que a matéria de quarks \textit{ud} livres~\cite{fune2012,witten1984}, para uma gama de valores da pressão de sacola. Iremos obter, a partir da teoria da relatividade geral, as relações massa-raio e massa-pressão central, a fim de definirmos as propriedades das estrelas estranhas, como massa e raio típicos, além de averiguar a sua estabilidade.  Este estudo é similar ao desenvolvido por Kettner \textit{et al.} (1995), onde eles analisam a estabilidade das estrelas estranhas e das estrelas charmosas (constituídas pelos quarks \textit{u,d,s} e \textit{c}) de forma mais completa (e complexa), levando em conta as oscilações radiais da estrela e o modelo de sacola do MIT~\cite{kettner1995}.}

Inicialmente, vamos obter as equações de estado para quarks não massivos e, posteriormente, para quarks massivos. A aproximação de temperatura zero é válida, pois, passado algum tempo da sua criação, essas estrelas têm temperaturas desprezíveis na escala nuclear. Enquanto isso, a aproximação de quarks não massivos (limite ultrarrelativístico) é apropriada por estarmos tratando apenas de quarks leves~\cite{weber1999}. Por fim, faremos a comparação dos resultados obtidos para quarks não massivos e massivos (caso relativístico), analisando as regiões de estabilidade da estrela e os impactos que variações na pressão de sacola infligem nos resultados. Por simplicidade, consideraremos uma estrela estranha no final de sua evolução, implicando que esta estrela fria perdeu sua energia de rotação e está estática. Por ser estática e esfericamente simétrica, além de composta por um gás ideal relativístico e degenerado de quarks, é apropriado o uso da equação TOV.

Por ser o sistema de unidades mais apropriado em Física de Partículas, utilizaremos as chamadas unidades naturais, onde $\hbar=c=k_B=\num{1}$. Para mais detalhes, apresentamos no Apêndice A, do Material Suplementar, um comparativo com o sistema internacional de unidades (SI) e uma explicação mais detalhada sobre as unidades utilizadas neste trabalho.

\section{Descrição de uma estrela}
\label{sec:stardescription}
\begin{figure}[!tb]
  \centering
  \includegraphics[width=.45\textwidth]{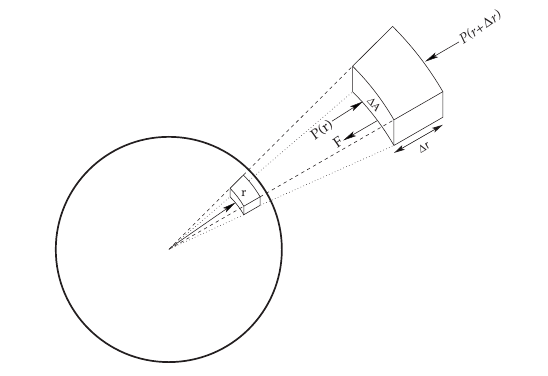}
  \caption{Ilustração esquemática das pressões atuando sobre um elemento infinitesimal de massa, a uma distância $r$ do centro da estrela~\cite{jackson2005}.}
  \label{fig:pressures}
\end{figure}

Uma estrela é definida como sendo uma esfera autogravitante em equilíbrio hidrostático, ou seja, que em sua superfície a pressão interna que tende a expandir a estrela é igual a pressão gravitacional que tende a comprimi-la, como mostra a Fig.~\ref{fig:pressures}. A descrição de uma estrela é feita por um conjunto de três equações: uma equação diferencial que descreve a variação da pressão perante o raio, outra equação diferencial correspondendo à variação da massa com o raio e uma terceira equação correspondente à pressão exercida pelo gás que constitui a estrela, chamada de \textbf{equação de estado}, que é obtida a partir da mecânica estatística.

A descrição gravitacional do ponto de vista newtoniano pode ser obtida da seguinte forma. Temos que a pressão sobre uma superfície esférica e a força correspondente são relacionadas simplesmente por
\begin{equation}
  \label{eq:dp}
  \mathrm{d}p = \frac{\mathrm{d}F}{4\pi r^2}\,,
\end{equation}
onde $p$ é a pressão, $F$ a força e $r$ a distância radial. {No equilíbrio, $\mathrm{d}F$ é compensada pela força gravitacional entre $m(r)$ e $\mathrm{d}m$} de tal forma que
\begin{equation}
  \label{eq:df}
  \mathrm{d}F=-\frac{G\, \mathrm{d}m\, m(r)}{r^2}\,,
\end{equation}
{onde $G$ é a constante gravitacional}. Por outro lado, o elemento infinitesimal de massa pode ser relacionado com a densidade de massa $\rho$ (como uma função de $r$), sendo dado por
\begin{equation}
  \label{eq:dm}
  \mathrm{d}m = \rho(r)\mathrm{d}V=\rho(r)4\pi r^2\mathrm{d}r\,,
\end{equation}
onde $V$ é o volume da esfera.

{A teoria da relatividade restrita implica que a densidade de massa $\rho$ pode ser expressa em termos da densidade de energia $\epsilon$ por $\epsilon = \rho c^2$.} Em unidades naturais (onde c = 1), elas são equivalentes. Substituindo a Eq.~(\ref{eq:dm}) na Eq.~(\ref{eq:df}), e o resultado na Eq.~(\ref{eq:dp}) obtemos
\begin{equation}
  \label{eq:dpdr}
  \frac{\mathrm{d}p}{\mathrm{d}r}=-\frac{G\epsilon(r)m(r)}{r^2}\,,
\end{equation}
que juntamente a
\begin{equation}
  \label{eq:dm1dr}
  \frac{\mathrm{d}m}{\mathrm{d}r}=4\pi r^2 \epsilon(r)\,,
\end{equation}
fornecem a descrição gravitacional de uma estrela newtoniana. A Fig.~\ref{fig:pressures} mostra a pressão do gás atuando em um elemento infinitesimal de massa, localizado a uma distância $r$ do centro da estrela, juntamente à força graviacional.  

Ao considerarmos o caso relativístico geral, a Eq.~(\ref{eq:dm1dr}) é a mesma. Mas a curvatura do espaço-tempo altera a relação entre a pressão e o raio para as estrelas mais densas, que geram um campo gravitacional mais intenso. A solução das equações de Einstein que descreve o comportamento da pressão no interior de uma estrela estática, esfericamente simétrica e composta por um fluido ideal é a chamada equação de Tolman-Oppenheimer-Volkoff (TOV)~\cite{oppenheimer1939,tolman1939}, cuja derivação é apresentada no Material Suplementar (Apêndice B), dada por
\begin{widetext}
  \begin{equation}
    \label{eq:tov}
    \frac{\mathrm{d}p}{\mathrm{d}r}=-\frac{G\epsilon(r)m(r)}{r^2}\left[1 + \frac{p(r)}{\epsilon(r)}\right]\left[1 + \frac{4\pi r^3p(r)}{m(r)}\right]\left[1 - \frac{2Gm(r)}{r}\right]^{-1}\,.
  \end{equation}
\end{widetext}
O primeiro termo é idêntico àquele presente na Eq.~(\ref{eq:dpdr}), enquanto os outros termos podem ser interpretados como correções relativísticas, que fortalecem a ação do campo gravitacional. Vale notar que quando o termo $2GM/R$ (onde $M$ é a massa total da estrela e o $R$ é raio da estrela) aproxima-se da unidade, os efeitos relativísticos são dominantes tal que a gravitação newtoniana torna-se insuficiente. Esta situação acontece para massas elevadas ou raios pequenos, indicando, de forma geral, altas densidades. De fato, o último termo indica que se $R=2GM$, o chamado \textbf{raio de Schwarzchild}, atinge-se a singularidade~\cite{weber1999}. No caso do Sol, este raio possui valor de aproximadamente \SI{3}{\km}, ou seja, ao comprimirmos a massa do Sol neste raio atingiremos a singularidade, formando um buraco negro. 

Para resolver estas duas equações diferenciais acopladas (seja o caso newtoniano ou o caso relativístico), precisamos de uma terceira equação, a chamada equação de estado, que descreve a pressão em termos da densidade de energia e que depende da composição da estrela. Além disso, condições de contorno são necessárias. No ponto central da estrela sabemos que não há massa, portanto $m(r=0)=0$. Não sabemos exatamente a pressão no ponto central, então estipulamos uma pressão para este ponto tal que $p(r=0)=p_0$. Na superfície da estrela, teremos envolvido toda sua massa, implicando em $m(r=R)=M$. Por outro lado, devido ao equilíbrio entre as pressões neste ponto (equilíbrio hidrostático), teremos $p(r=R)=0$~\cite{bandecchi2019}. 
 
A condição necessária para que uma estrela seja estável é que a derivada de sua massa total em relação à pressão central seja positiva~\cite{weber1999}, já que um aumento na massa (aumento na pressão gravitacional) estará implicando em um aumento da pressão central, representando uma situação de equilíbrio. Caso contrário, o sinal negativo da derivada implica que um aumento da massa vem com uma diminuição da pressão central, logo, a pressão gravitacional será maior, resultando no colapso da estrela. A massa total e a pressão central fazem parte das condições de contorno, sendo que a pressão central é estimada no início dos cálculos pela equação de estado. Enquanto isso, a massa total é obtida ao fim dos cálculos, quando integramos a equação de Newton, ou a equação TOV, perante todo o raio da estrela, satisfazendo a propriedade de que na superfície da estrela a pressão é nula, devido ao equilíbrio hidrostático. Para fins didáticos, os códigos em Fortran 90 utilizados para solucionar a equação TOV para as anãs brancas e para as estrelas estranhas estão disponíveis online~\footnote{https://github.com/llazzari/Codigos}.

\subsection{Anãs brancas}
\label{sec:wd}
Devido à elevada densidade presente nas anãs brancas, ficou claro que a pressão de radiação não era capaz de sustentar o colapso gravitacional. Em 1926, Fowler usou a estatística de Fermi-Dirac para explicar a sustentação de uma anã branca~\cite{fowler1926}. Fowler descreveu o gás que constitui uma anã branca como um gás de elétrons degenerados, que exercem a chamada pressão de degenerescência, que contrapõe a pressão gravitacional~\cite{jackson2005}. Essa pressão de degenerescência é consequência direta do princípio da exclusão de Pauli, onde dois férmions não podem ocupar o mesmo estado quântico simultaneamente~\cite{maciel1999}. No caso dos elétrons, a descrição do estado quântico é feita pelo \textit{spin} (\textit{up} ou \textit{down}) e pela energia. Desta forma, um gás degenerado possui pelo menos dois elétrons em cada nível de energia, ou seja, cada elétron com a mesma energia possui o \textit{spin} oposto do outro. 
{Portanto, em um gás de elétrons têm-se que a pressão de degenerescência leva os elétrons a ocuparem níveis de energia cada mais elevados. Considerando que estes estejam livres, teremos que a velocidade destes elétrons será cada vez maior. Entretanto, da teoria da relatividade restrita, sabemos que a velocidade da luz é a velocidade limite que as partículas do gás podem possuir. Portanto,  fica claro que deve haver uma massa limite que os elétrons sejam capazes de suportar. Utilizando-se destes conceitos,  Chandrasekhar  descobriu que a massa limite de uma anã branca era dada por~\cite{chandra1931}}
$$M_{\mathrm{Ch}} = 1,4312\left(\frac{2Z}{A}\right)^2\;\mathrm{M}_\odot\;,$$
onde $A$ é o número de núcleons ou número bariônico e $Z$ o número de prótons. Em sua homenagem, a massa limite é chamada de massa de Chandrasekhar~\cite{chandra1931,pinochet2016,sagert2006}. É possível visualizar que o aumento incondicional da pressão gravitacional (aumento da massa da estrela) não pode ser sempre sustentado por elétrons. A partir da $M_{\mathrm{Ch}}$, os elétrons não são mais capazes de suportar o colapso gravitacional, e outras partículas, como nêutrons e/ou quarks, devem desempenhar tal papel~\cite{weber1999}.

Como a massa dos elétrons é muito menor que a dos núcleons, eles são os primeiros a se tornarem degenerados, tornando a pressão dos íons e a de radiação negligenciáveis~\cite{sagert2006}. É a pressão de degenerescência que fornece oposição ao colapso gravitacional. Como consequência, o transporte de energia no interior dessa estrela se dá pela condutividade térmica dos elétrons. Isto é tão eficiente que o interior de uma anã branca é praticamente isotérmico, com queda na temperatura significativa apenas nas regiões externas não degeneradas~\cite{maciel1999}. Estas camadas externas radiam o calor restante, resfriando a anã branca lentamente, resultando em mudanças de cor da superfície, passando de branco para vermelho até se tornar um corpo negro. Conforme a anã branca esfria, o gás de elétrons vai se tornando cada vez mais degenerado. O tempo de resfriamento de uma anã branca é comparável com a idade do Universo, e as anãs brancas de brilho mais fraco servem como limites inferiores à idade do Universo~\cite{jackson2005}.     

Por serem \num{2000} vezes menos massivos que os núcleons, desprezaremos as contribuições dos elétrons para a massa, porém, eles fornecerão toda a pressão de degenerescência do gás, já que são as primeiras partículas a se tornarem degeneradas. Isto significa que também aproximaremos os núcleons como estando em repouso, contribuindo para a densidade de energia somente através de sua energia de repouso, sem contribuir para a pressão do gás. Outra aproximação que levaremos em conta, é a de que o gás se encontra em temperatura nula, o que é válido quando $(\mu - m_e) \gg T$, onde $\mu$ é o potencial químico, $m_e$ a massa do elétron e $T$ a temperatura, assim, o gás está completamente degenerado. Desta forma, segue que a pressão, a densidade de energia e a densidade de partículas são, respectivamente, dadas por~\cite{sagert2006}
\begin{widetext}
\begin{align}
  \label{eq:elecpress}
  p(x) = {} & \frac{m_e^4}{24\pi^2}[(2x^3-3x)(1+x^2)^{1/2}+3\sinh^{-1}x]\,, \\
  \epsilon(x) = {} & n\,m_N\frac{A}{Z} + \frac{m_e^4}{8\pi^2}[(2x^3+x)(1+x^2)^{1/2}-\sinh^{-1}x]\,, \\
  n = {} & \frac{k_F^3}{3\pi^2}\,,
\end{align}
\end{widetext}
onde $x=k_F/m_e$ e
\begin{equation}
  k_F=\left(\frac{3\pi^2\rho}{m_N}\frac{Z}{A}\right)^{1/3}
\end{equation}
é o momento de Fermi, valor limite do \textit{momentum} que os elétrons podem possuir. Além disso, $m_N$ é a massa de um núcleon. Portanto, no caso de uma anã branca composta por $^{12}$C e $^{16}$O, $Z/A = 1/2$.

\begin{figure}[!t]
  \centering
  \includegraphics[width=.5\textwidth]{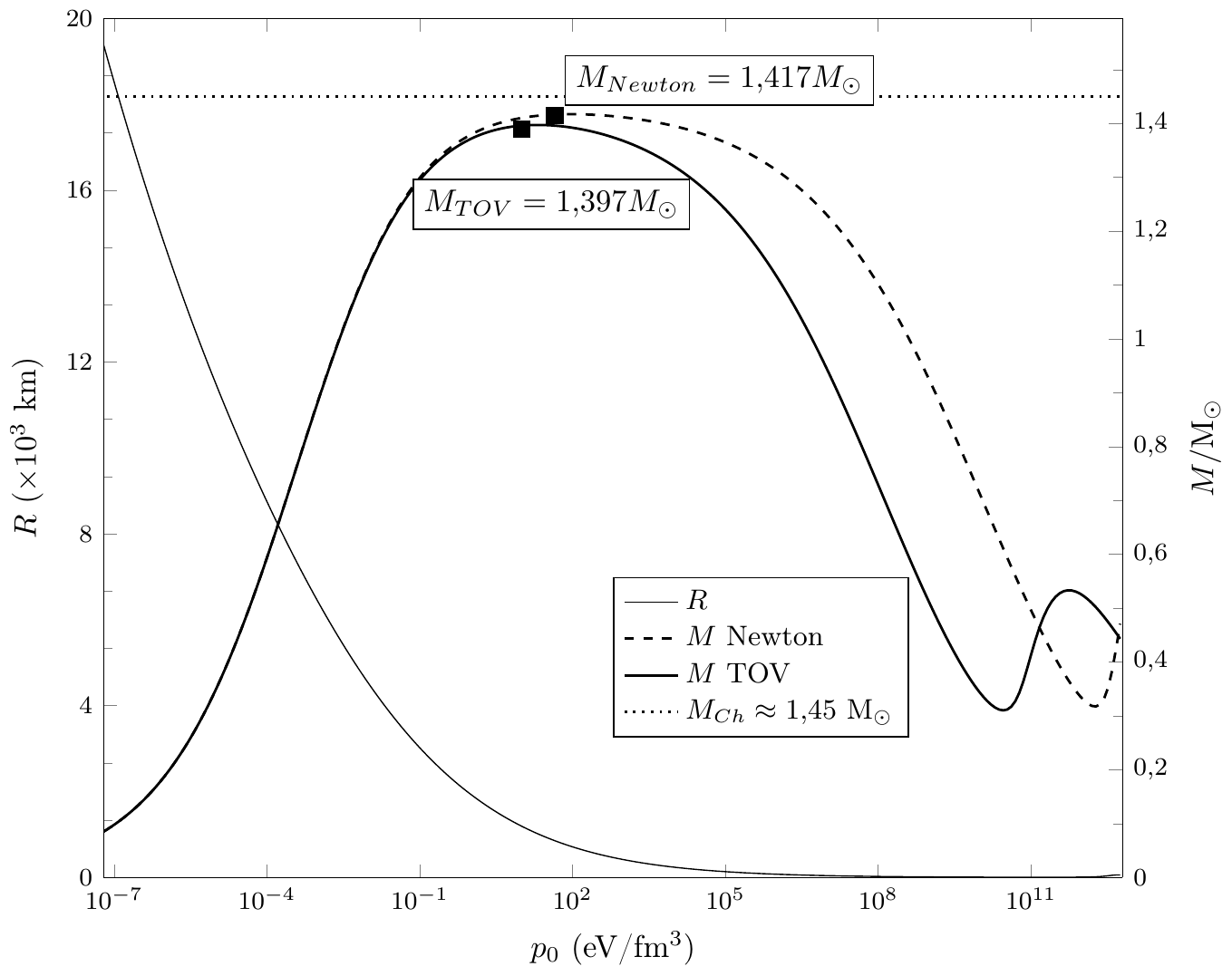}
  \caption{Relações massa-pressão central e raio-pressão central de uma estrela anã branca, sustentada pela pressão de degenerescência dos elétrons. Vale notar que este sistema é (a princípio) aceitável fisicamente, já que respeita a massa de Chandrasekhar.}
  \label{fig:wd}
\end{figure}

Anãs brancas são o único tipo de objeto compacto que pode ser tratado via gravitação newtoniana. Para uma anã branca típica com $M =$ \num{0,8}M$_{\odot}$ e $R =$ \SI{7000}{\km}, temos que $2GM/R\simeq$ \num{3e-4}, o que é muito menor do que a unidade, sendo assim, a aproximação newtoniana permanece válida. Porém, esta validade se estende até certo ponto. Para as anãs brancas mais densas, deve empregar-se a relatividade geral~\cite{sagert2006}, o que pode ser visto na Fig.~\ref{fig:wd}. Para obter estes resultados fizemos uso da equação de estado~[Eq.~(\ref{eq:elecpress})], além da equação da massa~[Eq.~(\ref{eq:dm1dr})], juntamente à descrição da pressão via gravitação newtoniana~[Eq.(~\ref{eq:dpdr})] e via relatividade geral, pela Equação TOV~[Eq.~(\ref{eq:tov})], a fim de compararmos as duas predições. Pode-se observar que o tratamento newtoniano das anãs brancas é válido para toda a região onde \num{0,1}\,M$_{\odot} \leq M \leq$ \num{1,4}\,M$_{\odot}$, implicando em \SI{1000}{\km} $\leq R \leq$ \SI{20000}{\km}. Esta análise mostra, também, que a massa máxima da anã branca é menor no caso relativístico do que no caso newtoniano. Isto ocorre devido aos termos adicionais da equação TOV, que fortalecem a intensidade do campo gravitacional, gerando para uma mesma massa um campo mais intenso do que aquela da gravitação de Newton. Além disso, pode-se observar que as soluções obtidas respeitam a massa de Chandrasekhar $M_{\mathrm{Ch}}$.  

A partir da Fig.~\ref{fig:wd}, também é possível analisar a estabilidade da estrela, através do sinal da derivada da sua massa total $M$ em relação à pressão central $p_0$. A região de estabilidade da estrela deve satisfazer a relação $dM/dp_0>0$, pois, se houver um incremento de sua massa e o mesmo ocorrer com a pressão central, o equilíbrio entre a pressão gravitacional e a pressão de degenerescência é mantido~\cite{sagert2006}. Por outro lado, se houver um incremento na massa da estrela, e a pressão central não se alterar, ou até mesmo, diminuir, a ação gravitacional levará ao colapso da estrela. Portanto, da Fig.~\ref{fig:wd}, temos duas regiões de estabilidade: a primeira para os valores aproximados de pressão central entre \SI{e-7}{eV/\femto\cubic\m} $\leq p_0 \leq$ \SI{10}{eV/\femto\cubic\m}, onde a massa varia aproximadamente entre \num{0,1}\,M$_{\odot} \leq M \leq$ \num{1,4}\,M$_{\odot}$; a segunda, corresponde ao intervalo de pressões centrais entre \SI{e10}{eV\per\femto\cubic\m} $\leq p_0 \leq$ \SI{e12}{eV\per\femto\cubic\meter}, e de pequenas massas entre \num{0,31}\,M$_{\odot}$ $\leq M \leq$ \num{0,55}\,M$_{\odot}$. A primeira região de estabilidade pode ser (e usualmente é) tratada via gravitação newtoniana em ótima aproximação. No segundo caso, a gravitação newtoniana leva a valores não aceitáveis e a teoria da relatividade geral deve ser empregada~\cite{sagert2006}.

\subsection{Estrelas de nêutrons}
\label{sec:ns}
No final de suas evoluções, estrelas massivas, com massas superiores a oito massas solares, expelem grandes quantidades de energia ao espaço interestelar, na forma de uma supernova. Antes disso, enquanto o núcleo de ferro da estrela original continua a contrair, os elétrons tornam-se relativísticos. Neste ponto, a estrela assemelha-se à uma anã branca, porém, contém elementos químicos mais pesados. Como a massa da estrela era, originalmente, elevada, a massa do núcleo de ferro será maior que a massa de Chandrasekhar, implicando que os elétrons não são capazes de suportar o colapso gravitacional. As altas energias cinéticas dos elétrons fazem com que seja favorável que ocorra o decaimento beta inverso, isto é, que prótons e elétrons combinem-se, formando nêutrons e neutrinos. Os neutrinos escapam da estrela, levando parte da energia gravitacional, mas com a densidade crescente, alguns ficam presos. O objeto denso resultante é denominado estrela de nêutrons.

O termo ``estrela de nêutrons'' surgiu a partir da descoberta do nêutron por Chadwick em 1932~\cite{chadwick1932}, o que levou Baade e Zwicky a proporem, dois anos depois, que uma estrela remanescente de uma supernova poderia ser sustentada pela pressão de degenerescência dos nêutrons~\cite{baade1934}. Em 1939, Tolman, Oppenheimer e Volkoff realizaram os primeiros cálculos com relação às estrelas de nêutrons~\cite{oppenheimer1939,tolman1939}. Eles estimaram que o raio típico destes objetos seria de apenas \SI{10}{\km}. Além disto, neste estudo sobre estrelas de nêutrons, a partir das equações de campo de Einstein, Volkoff, sob a orientação de Oppenheimer e utilizando o livro de Tolman, deduziu a notável equação TOV~\cite{oppenheimer1939,tolman1939}. Anos depois, estimou-se que a conservação do fluxo magnético de uma gigante vermelha que é contraída gravitacionalmente em uma estrela de nêutrons, poderia produzir campos magnéticos da ordem \SI{e12}{G}! Por conta do seu pequeno raio, a luminosidade das estrelas de nêutrons foi estimada em valores muito baixos, sendo considerado na época, impossível de serem detectadas.

Pelas observações realizadas, estava claro que um pulsar deveria ser uma estrela compacta, sendo que o único tipo observado até então, eram as anãs brancas~\cite{weber1999}. Em 1968, Gold propôs que os pulsares detectados seriam estrelas de nêutrons girando rapidamente, com elevados campos magnéticos~\cite{gold1968}. Dentro da explicação de Gold para associar um pulsar a uma estrela de nêutrons, deve-se interpretar que a pulsação de um pulsar é causada pelo período de rotação, e não por sua vibração. Ele argumentou que a vasta quantidade de energia radiada, juntamente à enorme quantidade de energia rotacional armazenada em um objeto de aproximadamente \SI{10}{\km} e com um fluxo magnético tão grande, ambos conservados da estrela progenitora, sugerem que tal objeto girando rapidamente poderia explicar um período estável de emissão de sinais, como aqueles observados. Além destes argumentos, a detecção de pulsares via radiação eletromagnética implica que estes objetos estão perdendo energia. Desta forma, a amplitude de vibração diminui com o passar do tempo, o que não afeta a frequência observada.
No caso da rotação associada à frequência de emissão, esta é amortecida suavemente o que acarreta em uma mudança acumulada no período de rotação, de três segundos em \num{e8} anos~\cite{weber1999}, estando de acordo com os resultados observacionais.

Nos trabalhos de Oppenheimer e Volkoff foi estimado, com uma equação de estado similar à Eq.~(\ref{eq:elecpress}) e dada por~\cite{sagert2006}
\begin{widetext}
  \begin{align}
    \label{eq:neutronpress}
    p(x)= {} &\frac{m_N^4}{24\pi^2}[(2x^3-3x)(1+x^2)^{1/2}+3\sinh^{-1}x]\,, \\
    \epsilon(x) = {} & n\,m_N + \frac{m_N^4}{8\pi^2}[(2x^3+x)(1+x^2)^{1/2}-\sinh^{-1}x]\,, \\
    n = {} & \frac{k_F^3}{3\pi^2}\;,
  \end{align}
\end{widetext}

onde $x=k_F/m_N$ e $$k_F=\left(\frac{3\pi^2\rho}{m_N}\right)^{1/3}$$ é o momento de Fermi; que um gás de nêutrons livres não seria capaz de sustentar estrelas com massas superiores a \num{0,7}\,M$_\odot$. Desta forma, ficou claro que estrelas de nêutrons são compostas por outras partículas, além de nêutrons~\cite{weber1999}.

As evidências sugerem que possamos dividir uma estrela de nêutrons em quatro camadas~\cite{weber1999}. A camada mais externa é chamada de atmosfera, possui poucos centímetros e é composta por átomos de hidrogênio e hélio. A camada logo abaixo, chamada de crosta externa, é composta por uma rede de núcleos atômicos, além de um gás degenerado e relativístico de elétrons (essencialmente, a mesma matéria encontrada nas anãs brancas)~\cite{glendenning2007}. A crosta interna é composta por nêutrons em um estado superfluido. Por fim, no centro da estrela está o núcleo, onde todos os núcleos atômicos foram dissolvidos em seus constituintes, ou seja, em prótons e nêutrons. Como o regime de densidade presente é extremamente elevado, estados exóticos da matéria podem aparecer, desde píons e káons à híperons e quarks livres~\cite{nyiri2001}.

A descrição do núcleo da estrela de nêutrons está diretamente associada à teoria das interações fortes, a QCD. Há incertezas na descrição de sistemas com esse regime de densidade pela QCD, isto é, em sistemas onde a teoria perturbativa não é mais aplicável. Desta forma, o nome ``estrela de nêutrons'' leva ao conceito errôneo de uma estrela composta majoritariamente por nêutrons. Enquanto, na verdade, este nome abrange uma categoria vasta de diferentes objetos teóricos.

\begin{figure}[!t]
  \centering
  \includegraphics[width=.575\textwidth]{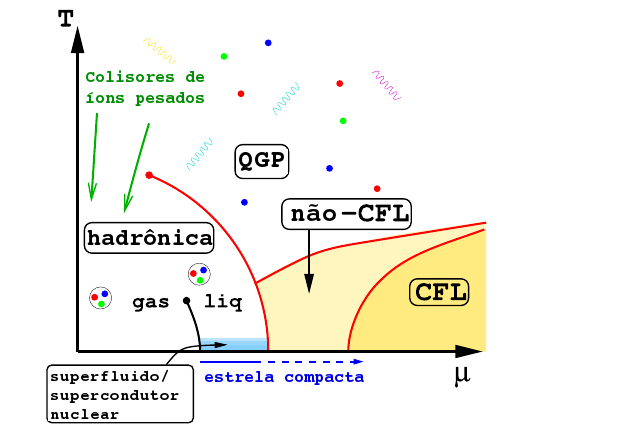}
  \caption{Diagrama representando as fases da matéria como função do potencial químico e da temperatura~\cite{alford2019}.}
  \label{fig:realphasetrans}
\end{figure}

A dificuldade em determinar qual o estado da matéria e seus constituintes no interior de um pulsar advém das elevadas densidades presentes nesses objetos, que estão muito além dos limites físicos testados em laboratórios~\cite{wong1994}. A fim de solucionar esta dificuldade, devemos compreender como funcionam as interações entre nêutrons, a possível formação de outros hádrons e, até mesmo, a transição de fase da matéria hadrônica para a matéria de quarks. Este último ponto, em especial, ganhou muita atenção nos últimos anos, e hoje, supomos que o diagrama correspondente à transição de fase seja da forma mostrada na Fig.~\ref{fig:realphasetrans}~\cite{alford2019}. Este diagrama relaciona as fases da matéria com a temperatura e com o potencial químico, lembrando que o potencial químico está associado à densidade bariônica. Desta forma, o diagrama mostra que a fase hadrônica pode aparecer nas formas de gás, líquido (matéria nuclear) e de superfluido. Fora da fase hadrônica, para baixas temperaturas, podemos ter uma matéria de quarks \textit{u, d} e \textit{s}, formando pares de Cooper em uma fase supercondutora de cor (CFL, do inglês \textit{color-flavor locking})~\cite{alford2019}. Em altas temperaturas e densidades (elevado potencial químico) chegamos ao plasma de quarks e glúons (QGP). {Atualmente, este novo estado da matéria é criado em colisões de íons pesados nos aceleradores de partículas RHIC e LHC \citep{Proceedings:2019drx}.} A transição de fase da matéria hadrônica para uma matéria de quarks livres permanece um tema de intenso estudo, devido à complexidade do diagrama apresentado na Fig.~\ref{fig:realphasetrans}.

\section{Estrelas estranhas}
\label{sec:strangestars}

\subsection{Cromodinâmica Quântica e o modelo de sacola do MIT}
\label{sec:qcd-mitbagmodel}

A teoria quântica de campos responsável pela interação forte é a QCD, onde a partícula mediadora da interação é o glúon, e a força se deve à carga de cor, presente apenas em quarks e glúons. Diferentemente do fóton, que media as interações eletromagnéticas e não possui carga elétrica, o glúon porta a carga associada à interação forte~\cite{wong1994}. Os quarks são os únicos férmions elementares que interagem através de todas as forças fundamentais, pois possuem as quatro cargas responsáveis pelas quatro interações fundamentais: cor, carga elétrica, sabor e massa. Desta forma, quarks são definidos a partir da sua carga de cor, carga elétrica, massa e spin. Os quarks possuem carga elétrica fracionária, sendo que os sabores \textit{u}, \textit{c} e \textit{t} possuem carga +2/3, enquanto os sabores \textit{d}, \textit{s} e \textit{b} têm carga -1/3. Além disso, os quarks \textit{d}, \textit{u} e \textit{s} são os ditos quarks leves, e possuem massas de \SI{0,003}{GeV}, \SI{0,005}{GeV} e \SI{0,1}{GeV}, respectivamente. Os quarks pesados, \textit{c}, \textit{b} e \textit{t}, possuem massas de \SI{1,3}{GeV}, \SI{4,5}{GeV} e \SI{174}{GeV}, respectivamente~\cite{silva2008}.

Existem diversas evidências experimentais acerca da existência de quarks. Porém, partículas de carga elétrica fracionária nunca foram detectadas diretamente, em outras palavras, quarks nunca foram detectados livres~\cite{silva2008}. A explicação para esta evidência experimental vem da hipótese do confinamento de cor, que postula que objetos que possuem carga de cor estão sempre confinados em estados com carga de cor nula. Sendo assim, somente estados incolores podem se propagar livremente na natureza. Acredita-se que o confinamento seja ocasionado pelas interações entre glúons, já que estes possuem a carga de cor e podem interagir entre si, restringindo esta força a curtos alcances.

Até hoje, provas analíticas do conceito de confinamento não foram obtidas. Entretanto, uma visão qualitativa pode ser obtida pelo exemplo de um quark e um antiquark livres sendo afastados continuamente. A interação entre eles será mediada por glúons virtuais que, por carregarem carga de cor, irão se atrair, comprimindo o campo de cor em um tubo entre os quarks. Em distâncias relativamente grandes, a densidade de energia do campo de glúons no tubo entre os quarks é constante. Sendo assim, a energia armazenada no campo de glúons será proporcional à distância entre os quarks~\cite{silva2008}. Este crescimento linear da energia armazenada com a distância, requer uma quantidade infinita de energia para separar o par quark-antiquark! Desta forma, torna-se energeticamente mais favorável a formação de um novo par quark-antiquark, ao longo do tubo, do que a separação dos quarks. A principal implicação é que apenas combinações de quarks e glúons com carga de cor nula podem existir nas escalas de energia (distâncias) presentes no nosso cotidiano.

Na QCD, a intensidade da interação depende das distâncias envolvidas. Sendo assim, quando a constante de acoplamento (medida da intensidade da interação) for pequena, um tratamento perturbativo é válido, caso contrário, o tratamento se torna não perturbativo. A descrição dos hádrons em escalas de baixas energias ou grandes distâncias espaciais, como as observadas em estrelas, não permitem o uso de teorias perturbativas, portanto, suas soluções são de difícil obtenção. A partir dos resultados obtidos pela QCD na rede~\cite{wong1994}, inspiraram-se modelos fenomenológicos, que permitem uma caracterização dos hádrons e de estados exóticos da matéria. 
{Como enfatizado na Introdução, o modelo mais simples  que reproduz os conceitos de confinamento e liberdade assintótica é o chamado de modelo de sacola do MIT~\cite{chodos1974}}. Nele, os hádrons são interpretados como uma região finita do espaço, isto é, uma ``sacola'', dentro da qual estão contidos os campos devidos aos quarks e glúons. É assumido no modelo que esta sacola é mantida por uma pressão constante $B$, chamada pressão de sacola, responsável por equilibrar a pressão exercida pelos quarks no interior da sacola, reproduzindo o conceito de confinamento. 

\begin{figure}[!t]
  \centering
  \includegraphics[width=.5\textwidth]{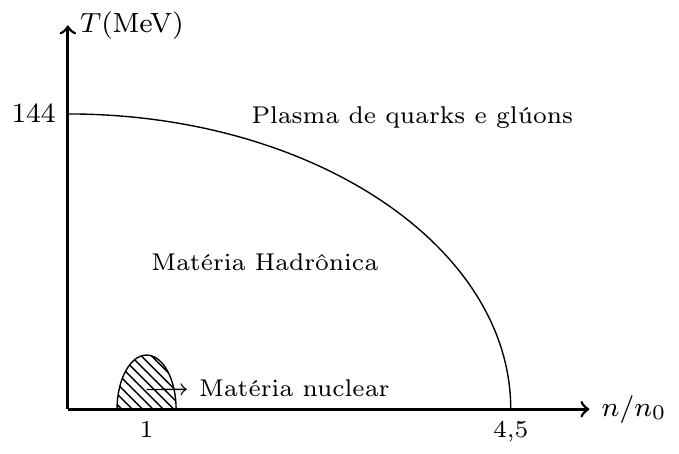}
  \caption{Diagrama de fase aproximado da QCD~\cite{wong1994}.}
  \label{fig:phase}
\end{figure}

Dentro deste modelo, são possíveis novas fases da matéria de quarks, já que se a pressão interna for maior do que a pressão de sacola, ocorrerá o rompimento do hádron e a matéria de quarks estará livre em uma região ``incolor''~\cite{weber1999}. Uma pressão cinética dos quarks, alta o suficiente para superar a pressão de sacola $B$, pode ocorrer quando a temperatura for alta e/ou a densidade bariônica for elevada, dando origem a um plasma de quarks e glúons. Por conta da complexidade da transição de fase apresentada na Fig.~\ref{fig:realphasetrans}, e do difícil tratamento relacionado à supercondutividade de cor na QCD, iremos assumir que a transição de fase da matéria hadrônica para a matéria de quarks e glúons livres pode ser descrita pelo diagrama de fase representado na Fig.~\ref{fig:phase}, o qual caracteriza uma transição de fase de primeira ordem. No Material Suplementar (Apêndice D) demonstramos que, para este diagrama de fase, a temperatura crítica a potencial bariônico nulo (considerando-se apenas dois sabores) é de aproximadamente \SI{144}{MeV}, para uma pressão de sacola $B^{1/4}=$ \SI{206}{MeV}~\cite{wong1994}. Tal valor de temperatura crítica é inferior àquele obtido usando QCD na rede para o regime de altas temperaturas e potencial bariônico nulo \citep{Borsanyi:2010bp}. Estes resultados também indicam que neste regime a transição de fase não é de primeira ordem, sendo sua descrição um tema de debate.
Por outro lado, a razão pela qual uma densidade elevada pode romper a sacola é, novamente, a pressão de degenerescência. A densidade bariônica da matéria ordinária é $n_0=$ \SI{0,16}{\per\cubic\femto\meter} e, para que ocorra o rompimento do hádron à temperatura zero e com $B^{1/4}=$ \SI{206}{MeV}, a densidade crítica é pelo menos \num{4,5}\,$n_0$~\cite{wong1994}. Densidades desta ordem podem estar presentes nas estrelas catalogadas como estrelas de nêutrons, ou pelo menos, no núcleo de algumas dessas, levando às diferentes possíveis formações destes objetos~\cite{weber1999}.

\begin{figure}[!t]
  \centering
  \includegraphics[width=.45\textwidth]{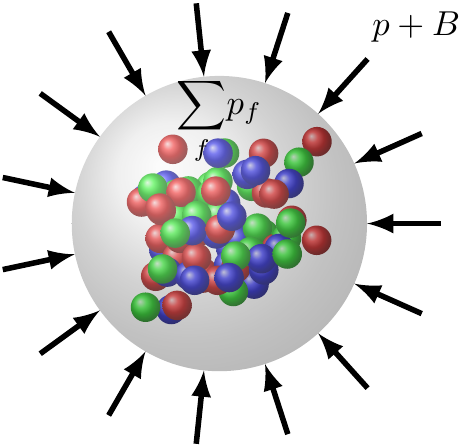}  
  \caption{Ilustração do equilíbrio das pressões para uma matéria de quarks livres em equilíbrio hidrostático~\cite{fune2012}.}
  \label{fig:bagpressure}
\end{figure}

Considerando que as densidades presentes nas estrelas de nêutrons sejam suficientes para formar uma matéria de quarks livres, é possível estender o modelo de sacola do MIT para descrever uma estrela composta por quarks livres. De fato, pode-se usar o modelo para estrelas em equilíbrio hidrostático, a fim de obter a equação de estado~\cite{weber1999}. Neste sentido, a estrela inteira é interpretada como uma sacola incolor, constituída por quarks livres. Ao considerarmos que a estrela possui simetria esférica, é estática e composta por um fluido ideal isotrópico, temos que a pressão no interior da estrela vai ser dada pela contribuição de cada sabor de quark à pressão cinética, identificada por $p_f$, onde o subíndice $f$ representa o sabor. Logo, pelo modelo de sacola do MIT, a pressão total $p$ é dada por~\cite{fune2012}
\begin{equation}
\label{eq:p1}
p = \sum_{f}p_f-B\;.
\end{equation}
Ilustramos o equilíbrio entre essas pressões para uma estrela densa composta por quarks livres na Fig.~\ref{fig:bagpressure}.

Como quarks são férmions, deve-se utilizar a estatística de Fermi-Dirac, da mesma forma que foi feito nas seções~\ref{sec:wd} e~\ref{sec:ns}, para obter uma equação de estado. Porém, dentro do modelo de sacola do MIT, deve-se levar em conta a pressão de sacola $B$, como mostra a Eq.~(\ref{eq:p1}). A partir disto, considerando que o fluido ideal de quarks que constitui a estrela é, na verdade, um gás ideal de Fermi completamente degenerado e ultrarrelativístico, tem-se que a pressão, a densidade de energia e a densidade bariônica em unidades naturais são, respectivamente, dadas por~\cite{weber1999}
\begin{align}
  \label{eq:pm0}
  p = {} & - B + \frac{1}{4\pi^2}\sum_f\mu_f^4 \,, \\
  \label{eq:epsm0}
  \epsilon = {} & B + \frac{3}{4\pi^2}\sum_f\mu_f^4 = 4B + 3p\,, \\
  \label{eq:rhom0}
  n = {} & \sum_f\frac{\mu_f^3}{3\pi^2}\,,
\end{align}
para um gás de quarks não massivos a temperatura zero, onde $\mu_f$ é o potencial químico para o quark de sabor $f$. Este limite de temperatura zero é aplicável para cálculos relativos às estrelas similares as estrelas de nêutrons, já que, pouco tempo após o seu surgimento, suas temperaturas caem para valores da ordem de \si{KeV}, o que é desprezível na escala nuclear. Para fins de comparação, um gás de quarks massivos à temperatura zero, possui a seguinte equação de estado
\begin{widetext}
  \begin{align}
    \label{eq:pT0}
    p = {} & -B + \sum_f\frac{1}{4\pi^2}\left[\mu_fk_f\left(\mu_f^2-\frac{5}{2}m_f^2\right)+\frac{3}{2}m_f^4\ln\left(\frac{\mu_f+k_f}{m_f}\right)\right]\,, \\
    \label{eq:eT0}
    \epsilon = {} & B + \sum_f\frac{3}{4\pi^2}\left[\mu_fk_f\left(\mu_f^2-\frac{1}{2}m_f^2\right)-\frac{1}{2}m_f^4\ln\left(\frac{\mu_f+k_f}{m_f}\right)\right]\,, \\
    \label{eq:rhoT0}
    n = {} & \sum_f\frac{k_f^3}{3\pi^2}\,,
  \end{align}
\end{widetext}
onde $k_f$ é o momento de Fermi associado ao quark de sabor $f$, definido em termos do potencial químico, já que $\mu_f^2=m_f^2+k_f^2$~\cite{weber1999}, onde $m_f$ é a massa do quark de sabor $f$.

\subsection{Hipótese de Bodmer-Witten-Terazawa}
\label{sec:bwt-hypothesis}
As equações de estado obtidas pelo modelo de sacola do MIT podem ser aplicadas para qualquer matéria composta por um gás de quarks livres, independente dos sabores de quarks sendo levados em consideração. Porém, quais sabores de quarks devem fazer parte de uma matéria formado por quarks livres? A hipótese de Bodmer-Witten-Terazawa~\cite{bodmer1971,terazawa1989,witten1984} consiste na afirmação de que a matéria estranha de quarks (SQM) é o verdadeiro estado {fundamental} da matéria que interage fortemente. Sendo assim, a SQM é mais estável do que a matéria nuclear ordinária e deve se fazer presente no interior da estrela. A fim de analisarmos a estabilidade da SQM, devemos comparar a sua energia de ligação -- densidade de energia superficial por densidade bariônica -- com a do isótopo $^{\footnotesize \num{56}}$Fe (o elemento mais estável encontrado na natureza), que é \SI{930}{MeV}~\cite{weber1999}. Desta forma, sabendo que o estado mais estável é aquele que representa a configuração de menor energia de ligação, a SQM deve satisfazer a condição
\begin{equation}
  \label{eq:estabilidadesqm}
  \frac{E}{A}\bigg{\vert}_{\mathrm{SQM}} < \frac{E}{A}\bigg{\vert}_{^{\num{56}}\mathrm{Fe}}\;,
\end{equation}
onde $E = V\epsilon_{\mathrm{sup}}$ é a energia de ligação, $\epsilon_{\mathrm{sup}}$ é a densidade de energia superficial e $V$ o volume ocupado pelo gás. Além disso, $A = Vn$ é o número bariônico, ou seja, $E/A=\epsilon_{\mathrm{sup}}/n$. É razoável assumir que a SQM faça parte de uma estrela com tamanha densidade, seja uma estrela híbrida ou uma estrela de quarks, pois espera-se que híperons -- hádrons que contenham o quark \textit{s} -- sejam produzidos pelas interações fortes entre nêutrons~\cite{weber1999}.

A partir do rompimento de nêutrons no interior da estrela, uma matéria de quarks \textit{u} e \textit{d} livres poderia ser formada. Entretanto, estes quarks estão presentes na matéria ordinária, e não parecem apresentar uma configuração de menor energia do que o isótopo $^{\footnotesize \num{56}}$Fe. Por esta razão, iremos impor a seguinte condição 
\begin{equation}
  \label{eq:estabilidadefe}
  \frac{E}{A}\bigg{\vert}_{^{\num{56}}\mathrm{Fe}} < \frac{E}{A}\bigg{\vert}_{\mathrm{u,d}}\,,
\end{equation}
que será utilizada para estipularmos os intervalos possíveis da pressão de sacola $B$.

Considerando as equações para quarks não massivos, Eqs.~(\ref{eq:pm0}),~(\ref{eq:epsm0}) e~(\ref{eq:rhom0}), na superfície da sacola (onde $p=0$) temos que a densidade de energia superficial é $\epsilon_{\mathrm{sup}}=4B$. Por outro lado, a pressão de sacola será dada por~\cite{weber1999}
\begin{equation}
  \label{eq:b}
  B = \frac{1}{4\pi^2}\sum_f\mu_f^4\;.
\end{equation}
Por simplicidade, iremos assumir a estrela como sendo eletricamente neutra, e descartaremos a presença de léptons carregados e neutrinos no interior da estrela. A condição de neutralidade elétrica global é dada por
\begin{equation}
  \label{eq:charge}
  q = \sum_fQ_fn_f = \frac{1}{3\pi^2}\sum_fQ_f\mu_f^3 =0\;,
\end{equation}
onde $q$ é a densidade de carga elétrica e $Q_f$ é carga elétrica do quark de sabor $f$. Desta forma, podemos relacionar os potenciais químicos dos quarks presentes na estrela.

A fim de verificarmos a condição representada na Eq.~(\ref{eq:estabilidadefe}), analisaremos primeiro pela condição de neutralidade global as relações entre os potenciais químicos para os quarks \textit{u} e \textit{d}. A partir da Eq.~(\ref{eq:charge}), temos que
\begin{equation}
  \label{eq:mu22} 
  \mu_d=2^{1/3}\mu_u\equiv \mu_2\,.
\end{equation}
Ao aplicarmos este resultado na Eq.~(\ref{eq:rhom0}), obtemos
\begin{equation}
  \label{eq:rhomu2}
  n = \mu_2^3/\pi^2\,.
\end{equation}
Além disso, usando a Eq.~(\ref{eq:b}) e isolando $\mu_2$ temos
\begin{equation}
  \label{eq:mu2}
  \mu_2= \left(\frac{4\pi^2}{1+2^{4/3}}\right)^{1/4}B^{1/4}\;.
\end{equation}
Logo, a energia de ligação, para uma matéria de quarks $ud$ livres, é tal que~\cite{nyiri2001}
\begin{widetext}
  \begin{equation}
    \label{eq:ud}
    \frac{E}{A}\bigg{\vert}_{\mathrm{u,d}}=\frac{4B\pi^2}{\mu_2^3}=(2\pi)^{1/2}(1+2^{4/3})^{3/4}B^{1/4} \simeq \SI{934}{MeV}\;,
  \end{equation}
\end{widetext}
onde usamos $B^{1/4}=$ \SI{145}{MeV}. A partir da comparação entre os valores da energia de ligação da matéria de quarks \textit{u} e \textit{d} livres e do isótopo $^{\footnotesize \num{56}}$Fe, vemos que o valor mínimo de $B^{1/4}$ deve estar em torno de \SI{145}{MeV}, já que para valores menores do que este, a matéria de quarks \textit{u} e \textit{d} livres seria mais estável do que a matéria ordinária, deixando de satisfazer a condição dada pela Eq.~(\ref{eq:estabilidadefe}).

No caso da SQM, temos que um número igual de quarks \textit{u, d} e \textit{s} satisfaz a condição de neutralidade elétrica~(\ref{eq:charge}), implicando em
\begin{equation}
  \label{eq:mu3mu}
  \mu_3\equiv\mu_u=\mu_d=\mu_s\,.
\end{equation}
Logo
\begin{equation}
  \label{eq:rho3}
  n = \mu_3^3/\pi^2\,.
\end{equation}
Desta forma, a partir da Eq.~(\ref{eq:b}), ao isolarmos $\mu_3$ obtemos
\begin{equation}
  \label{eq:mu3}
  \mu_3=\left(\frac{4\pi^2}{3}\right)^{1/4}B^{1/4}\,.
\end{equation}
Isto implica que a energia de ligação da SQM é
\begin{equation}
  \label{eq:ud}
  \frac{E}{A}\bigg{\vert}_{\mathrm{SQM}}=\frac{4B\pi^2}{\mu_3^3}=(2\pi)^{1/2}3^{3/4}B^{1/4} \simeq \SI{829}{MeV}\,.
\end{equation}

\begin{figure}[!t]
  \centering  
  \includegraphics[width=.5\textwidth]{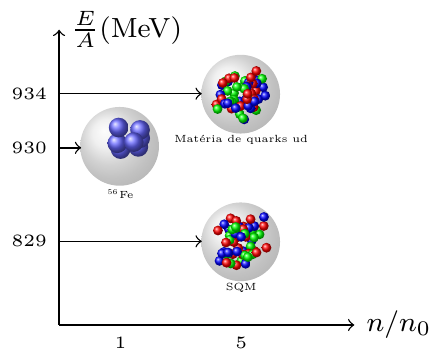}
  \caption{Comparação das energias de ligação por número bariônico em função da razão entre as densidades bariônica e a nuclear ($n_0=$ \SI{0,16}{fm^{-3}}) para análise da estabilidade dos tipos de matéria~\cite{weber1999}.}
  \label{fig:sqmstability}
\end{figure}

Através desta análise, provamos  que {o modelo de sacola do MIT implica que} a SQM é mais estável do que uma matéria contendo apenas quarks \textit{u} e \textit{d} livres e do que a matéria nuclear ordinária (lembrando que a energia de ligação do isótopo $^{\footnotesize \num{56}}$Fe é \SI{930}{MeV}). Sendo assim, concluímos que a hipótese de Bodmer-Witten-Terazawa é válida, como ilustramos na Fig.~\ref{fig:sqmstability}, para certos valores de $B$. Na verdade, a SQM é absolutamente estável em relação ao isótopo $^{\footnotesize \num{56}}$Fe se $B^{1/4}<$ \SI{162,8}{MeV}~\cite{weber1999} e, por esta razão, focaremos nosso trabalho nas estrelas estranhas.

{No modelo de sacola do MIT, a} estabilidade da SQM está diretamente associada aos valores da pressão de sacola $B$. Teoricamente, a SQM é completamente estável em relação ao isótopo $^{\footnotesize \num{56}}$Fe para $B^{1/4} < \SI{162,8}{MeV}$ e metaestável em relação a um gás de nêutrons para $B^{1/4}< \SI{164,5}{MeV}$~\cite{weber1999}. Esses são os limites superiores de $B$, enquanto que o limite inferior foi estabelecido pela condição dada na Eq.~(\ref{eq:estabilidadefe}) como sendo $B^{1/4} > \SI{145}{MeV}$. A análise dos valores de $B$, sejam experimentais, observacionais ou simulações de computadores, não são triviais. 

{A validade da hipótese de
Bodmer-Witten-Terazawa para modelos mais sofiscados para o tratamento das propriedades básicas da QCD do que  o modelo de sacola do MIT ainda é um tema de debate na literatura \citep{Klaehn:2017mux}. Entretanto, deve-se }ressaltar o fato de que a presença da matéria composta por átomos não contradiz a hipótese de Bodmer-Witten-Terazawa, de que a SQM é o estado fundamental da matéria que interage fortemente. Isto se deve ao fato que a matéria hadrônica, perante interações fracas, é estável por mais de \num{e60} anos! Um tempo muito maior do que a idade estimada do Universo de 13 bilhões de anos~\cite{bodmer1971}. Sendo assim, podemos esperar que a SQM só se faça presente no interior das estrelas mais densas e compactas, abrindo uma nova categoria para tais objetos. Por conta disto, esperamos que diferentes objetos possam ser formados nestas condições extremas de densidade. Nos casos menos densos, é consensual que as estrelas sejam compostas majoritariamente por nêutrons, enquanto configurações mais densas podem apresentar estados hadrônicos exóticos e até mesmo a SQM~\cite{weber1999}.

\subsection{Um exemplo simples}
\label{sec:results}
A fim de exemplificar os conceitos descritos acima, iremos no que segue apresentar um modelo simples de estrela estranha. Consideraremos as estrelas estranhas como estáticas, esfericamente simétricas, compostas pela SQM e sem crosta. Neste caso, as propriedades desta estrela podem ser obtidas a partir da solução da TOV~[Eq.(\ref{eq:tov})], juntamente a equação da massa~[Eq.~(\ref{eq:dm1dr})], assumindo a equações de estado derivadas a partir do modelo de sacola do MIT [Eqs.~(\ref{eq:pm0}) e~(\ref{eq:pT0})]. Compararemos os resultados para uma equação de estado que desconsidera a massa dos quarks (limite ultrarrelativístico) com uma equação de estado onde a massa dos quarks é levada em conta (caso relativístico), ambas à temperatura zero. Neste modelo, para um valor da pressão de sacola $B$, que satisfaça a hipótese de Bodmer-Witten-Terazawa, assumiremos a densidade central $n_c$ como condição inicial.

\begin{figure}[!t]
  \centering
  \includegraphics[width=.5\textwidth]{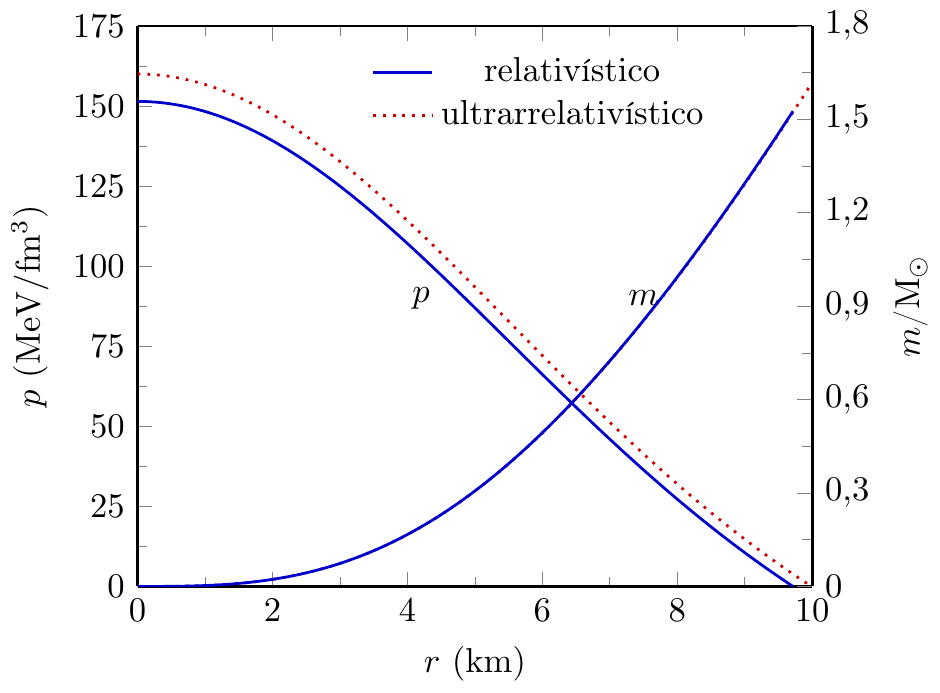}
  \caption{Pressão e massa em função do raio, no interior de uma estrela estranha com densidade bariônica central $n_c = \num{5}\,n_0$ e pressão de sacola $B^{1/4} = \SI{155}{MeV}$.}
  \label{fig:pmxr}
\end{figure}

Na Fig.~\ref{fig:pmxr}, mostramos o comportamento da pressão e da massa com o raio, do centro da estrela estranha até sua superfície, para $n_c = \num{5}\,n_0$ (onde $n_0 = \SI{0,16}{\femto\cubic\meter}$ é a densidade nuclear) e $B^{1/4}=\SI{155}{MeV}$. Na superfície da estrela, a pressão deve ser nula, por causa do equilíbrio hidrostático, ou seja, as pressões de degenerescência, gravitacional e de sacola se anulam neste ponto. Além disso, neste ponto atingimos a massa e o raio totais da estrela, que são, respectivamente, $M = \num{1,526}\,$M$_\odot$ e $R = \SI{9,715}{\km}$, no caso relativístico. Enquanto isso, no limite ultrarrelativístico obtemos $M = \num{1,615}\,$M$_\odot$ e $R = \SI{9,999}{\km}$. Apesar de auxiliar no entendimento acerca do comportamento da pressão e da massa a cada camada esférica concêntrica, este resultado tem pouca utilidade prática, já que representa uma única configuração de estrela estranha. 

A principal diferença entre os dois resultados da Fig.~\ref{fig:pmxr} é que as pressões centrais são distintas. Isto ocorre porque a relação entre pressão e densidade é dada de maneira distinta nos casos relativístico e ultrarrelativístico. Entretanto, a diferença nos resultados é, de fato, pequena como já fora apontada por Alcock \textit{et al.} na Ref.~\cite{alcock1986} e está em torno de 4\%. Isto ocorre porque estamos tratando de quarks leves, e suas massas exercem pouca influência na equação de estado e, consequentemente, nos parâmetros gerais da estrela. Sendo assim, a aproximação ultrarrelativística, extremamente utilizada na literatura, é bastante apropriada no estudo sobre estrelas estranhas.

\begin{figure}[!t]
  \centering
  \includegraphics[width=\columnwidth]{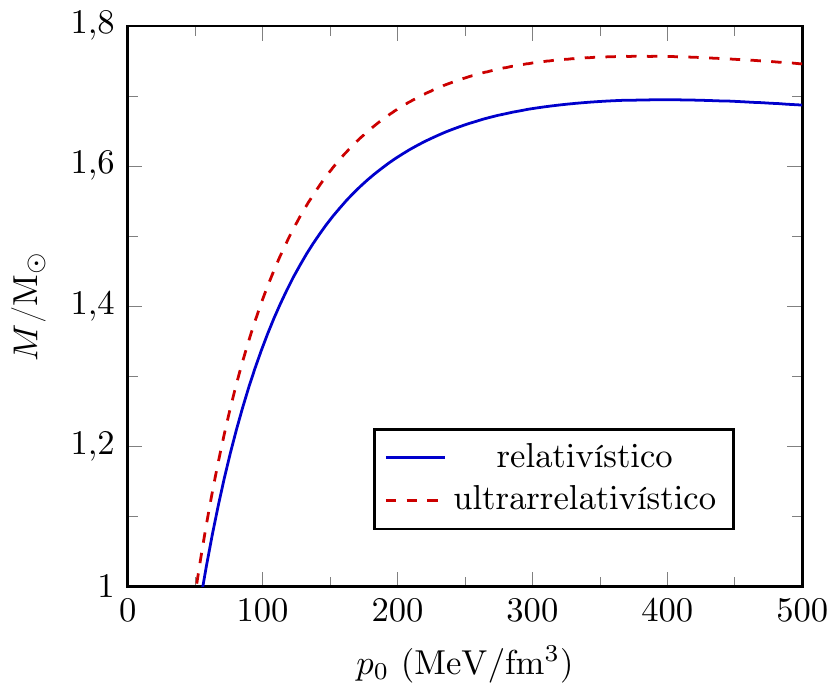}
  \caption{Perfil massa-pressão central nos limites relativístico e ultrarrelativístico. Cada ponto representa uma possível configuração de estrela estranha, para o mesmo valor da pressão de sacola $B^{1/4}=\SI{155}{MeV}$, e para densidades centrais entre \num{2}\,$n_0$ e \num{10}\,$n_0$.}
  \label{fig:Mxp0}
\end{figure}

\begin{figure}[!t]
  \centering
  \includegraphics[width=\columnwidth]{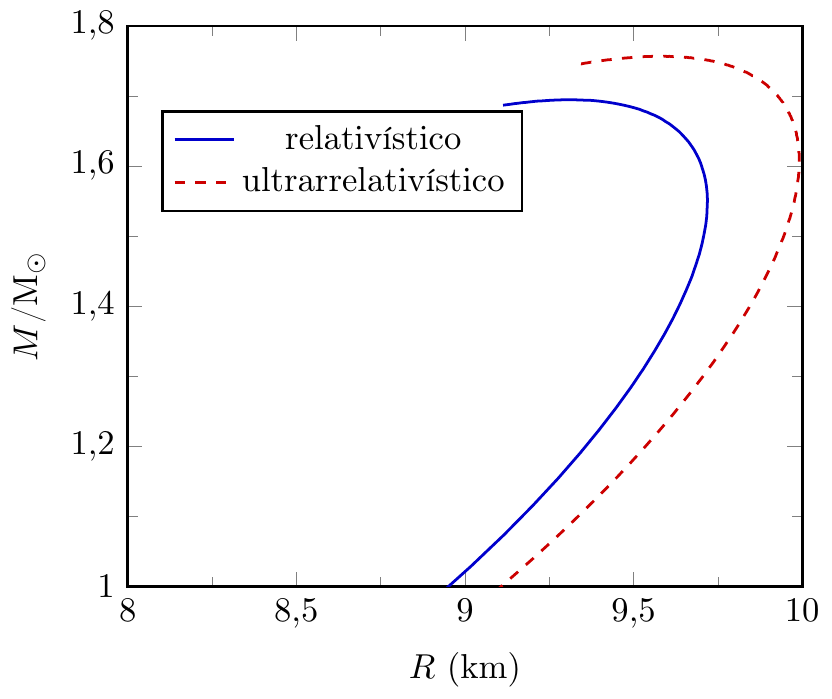}
  \caption{Perfil massa-raio nos limites relativístico e ultrarrelativístico, considerando $B^{1/4} = \SI{155}{MeV}$ e densidades centrais entre \num{2}\,$n_0$ e \num{10}\,$n_0$.}
  \label{fig:MxR}
\end{figure}

A fim de obtermos resultados que possam ser comparáveis com as observações de pulsares, devemos averiguar os resultados obtidos para a massa e raio totais, considerando cada valor de $p_0$. O processo de integração é o mesmo realizado anteriormente, mas agora estamos interessados nos valores finais que correspondem à densidade central predeterminada. Sendo assim, cada ponto nas figuras~\ref{fig:Mxp0} e~\ref{fig:MxR} caracteriza uma configuração possível, ou seja, determina para um dado valor de $n_c$, a massa e o raio da estrela. Todavia, o valor de $B^{1/4} = \SI{155}{MeV}$ é mantido nos dois casos.

\begin{figure*}[!t]
  \centering
  \includegraphics[width=\textwidth]{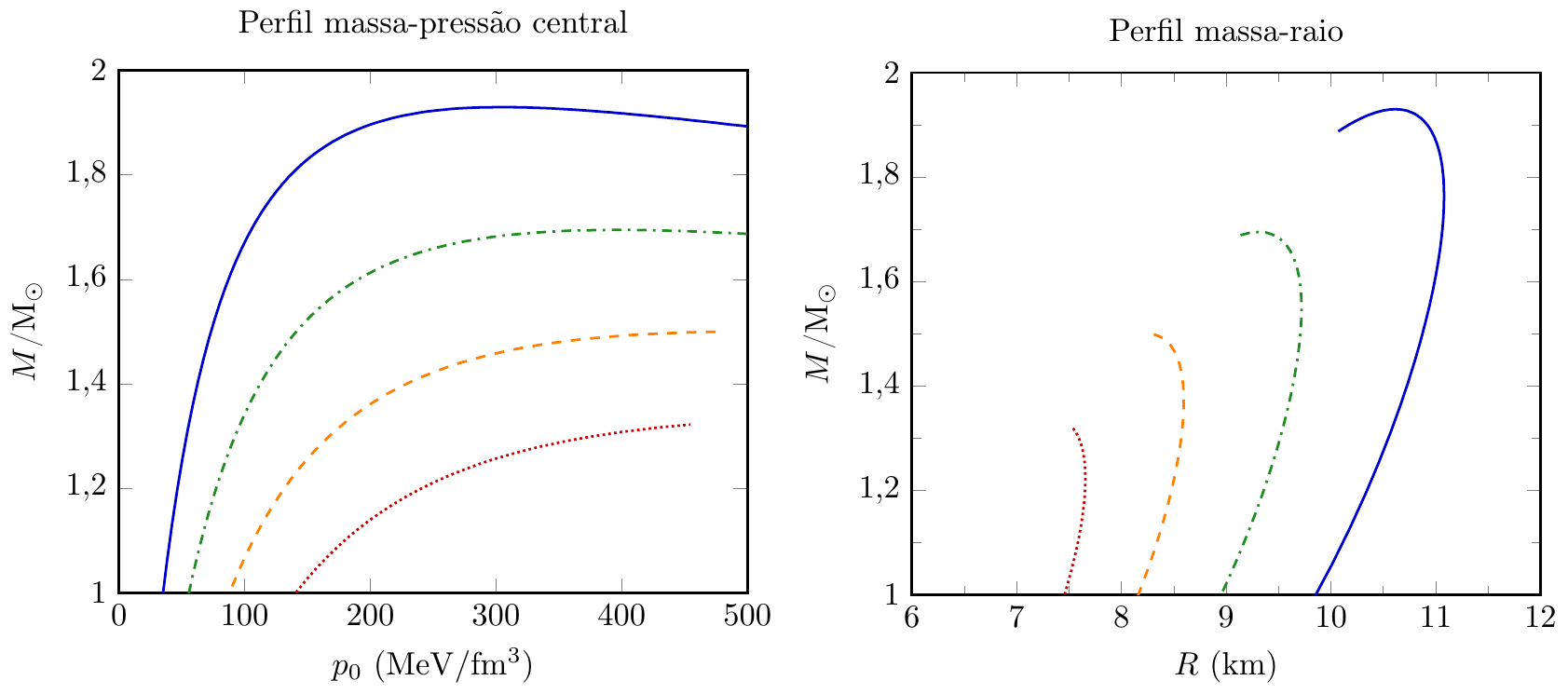}
  \caption{Perfis massa-pressão central e massa-raio, respectivamente, no caso relativístico, para os seguintes valores de $B$: linha sólida azul ($B^{1/4} = \SI{145}{MeV}$), linha traço-ponto verde ($B^{1/4} = \SI{155}{MeV}$), linha tracejada laranja ($B^{1/4}=\SI{165}{MeV}$) e linha pontilhada vermelha ($B^{1/4} = \SI{175}{MeV}$).}
  \label{fig:BMx}
\end{figure*}

Da Fig.~\ref{fig:Mxp0} podemos analisar a estabilidade das estrelas estranhas, pela condição apresentada na Seção~\ref{sec:stardescription}, onde $\mathrm{d}M/\mathrm{d}p_0 > 0$ implica em uma configuração estável. Desta forma, a região de estabilidade de uma estrela estranha, onde o crescimento da pressão central corresponde a um crescimento na massa total da estrela, é tal que $p_0 \leq \SI{396,35}{MeV\per\femto\cubic\meter}$, levando a uma massa máxima de \num{1,695}\,M$_\odot$ para $n_c = \num{8,598}\,n_0$, no caso relativístico. No limite ultrarrelativístico, a região de estabilidade corresponde a $p_0 \leq \SI{385,00}{MeV\per\femto\cubic\meter}$, e uma massa máxima de \num{1,757}\,M$_\odot$ para $n_c = \num{8,260}\,n_0$. A massa máxima do caso ultrarrelativístico é maior devido ao fato de que quarks sem massa conseguem atingir velocidades maiores, por isso, exercem uma pressão de degenerescência maior e podem suportar pressões gravitacionais mais elevadas, ou equivalentemente, massas maiores.

Podemos apenas estimar qual é a pressão no interior de uma estrela, ou seja, não temos como observar e medir tal propriedade. Sendo assim, faz-se necessário um resultado em termos de massa e raio, que são os únicos parâmetros comparáveis com as observações de estrelas deste tipo, já que estamos considerando-as estáticas. Na Fig.~\ref{fig:MxR}, podemos ver o perfil massa-raio de estrelas estranhas obtidos numericamente, dentro de nosso modelo. Estes resultados são compatíveis e próximos aos valores típicos encontrados para pulsares, via observação astronômica~\cite{alford2019,aziz2019}. No caso ultrarrelativístico, obtemos que a configuração de massa máxima, $M = \num{1,757}\,$M$_\odot$, contida em um raio de apenas \SI{9,573}{\km}. Enquanto no caso relativístico, $M = \num{1,695}\,$M$_\odot$ e $R = \SI{9,312}{\km}$. Estes valores são bastante próximos e coerentes com os pulsares mais densos já observados~\cite{alford2019,aziz2019}. 

Um aspecto importante na análise de estrelas estranhas é que as configurações onde $M < \num{1,0}\,$M$_\odot$ são ditas autoligadas~\cite{alcock1986}. Isto porque a força de ligação dominante é a força forte, e não a gravitacional. Em outras palavras, a pressão externa que mantém a estrela coesa é a pressão de sacola e não a pressão gravitacional. A atração gravitacional só começa a exercer influência sobre as possíveis configurações para massas maiores que uma massa solar. Por esta razão, nas figuras~\ref{fig:Mxp0},~\ref{fig:MxR} e~\ref{fig:BMx} só apresentamos os resultados para $M \geq \num{1,0}\,$M$_\odot$.

Variações nos valores da pressão de sacola $B$ alteram a estrutura e as possíveis configurações de uma estrela estranha composta por um gás de quarks massivos, como pode ser visto na Fig.~\ref{fig:BMx}. Esta variação também ocorre com quarks não massivos, porém focaremos no caso relativístico. O aumento em $B$ fortalece a contração da estrela e torna a equação de estado mais rígida, o que implica em massas máximas menores para as configurações correspondentes. Colocado de outra forma, o gás degenerado de quarks exerce a mesma pressão de degenerescência e deve lidar com uma pressão de sacola maior. Portanto, este consegue suportar uma pressão gravitacional menor. Outro efeito visível no aumento da pressão de sacola é a diminuição do raio, tornando a estrela estranha ainda mais compacta. Observações mais precisas nos observáveis -- massa e raio -- em pulsares candidatos a estrelas estranhas podem aumentar o conhecimento acerca da pressão de sacola e da transição de fase da matéria de quarks para a matéria hadrônica.   

O valor de $B$ escolhido para as figuras~\ref{fig:pmxr}, \ref{fig:Mxp0} e~\ref{fig:MxR} está dentro dos limites teóricos (\SI{145}{MeV} e \SI{162,8}{MeV}) para que a hipótese de Bodmer-Witten-Terazawa seja válida~\cite{weber1999} e que a SQM seja absolutamente estável em relação ao isótopo $^{\num{56}}$Fe.

\section{Considerações finais}
Neste trabalho apresentamos uma breve revisão das principais características e propriedades das estrelas compactas em geral, com foco nas estrelas estranhas. { Mostramos que em regimes de densidades elevadas, faz-se necessário o uso da teoria da relatividade geral para descrever  as anãs brancas, pois os resultados da teoria newtoniana divergem dos resultados relativísticos gerais para as pressões centrais mais elevadas.} Como as estrelas de nêutrons apresentam densidades ainda maiores, sua descrição está diretamente associada à solução da equação TOV. Devido às densidades presentes nas estrelas de nêutrons, sua constituição permanece um tema de estudo, {sendo possível o surgimento de estados exóticos em seu interior. Para fins didáticos, em nosso estudo consideramos que para altas densidades bariônicas, há uma transição de fase da matéria hadrônica para a matéria de quarks e que esta é  de  primeira ordem. Além disso, utilizamos o modelo de sacola do MIT para descrever as propriedades básicas da QCD. Dentro deste contexto demonstramos que a hipótese de Bodmer-Witten-Terazawa é válida para um dado intervalo de valores da pressão de sacola, o que implica na possibilidade da existência de estrelas estranhas, compostas inteiramente pela SQM}. Os resultados obtidos neste exemplo, permitiram ver que as estrelas estranhas {podem ser} estáveis e que sua estabilidade depende diretamente da equação de estado. A equação de estado, por sua vez, depende do parâmetro fenomenológico, a pressão de sacola, introduzida {de modo} \textit{ad hoc} no Modelo de Sacola do MIT. Concluímos, com esta simples abordagem, que estrelas estranhas {podem ser} uma das possíveis explicações para os pulsares mais densos observados.  {Por fim, enfatizamos que nosso objetivo foi apresentar os elementos básicos necessários à compreensão do tema. Consequentemente, foi necessário restringir a discussão sobre as inúmeras alternativas existentes para os pressupostos básicos assumidos em nossa análise, os quais são objeto de debate na literatura. Entretanto, acreditamos que o  modelo simplificado apresentado pode servir de base para que o estudante inicie seus estudos neste tema tão importante e atual em Astropartículas.}

\section*{Agradecimentos}
\label{thanks}
Este trabalho foi parcialmente financiado pela CNPq, FAPERGS e INCT-FNA (processo número 464898/2014-5). 

\section*{Material suplementar}
O seguinte material suplementar está disponível nos apêndices.
\begin{itemize}
\item[] Apêndice A: Sistema de unidades naturais.
\item[] Apêndice B: Derivação da equação de Tolman-Oppenheimer-Volkoff (TOV).
\item[] Apêndice C: Temperatura e potencial bariônico críticos para transição de fase de primeira ordem.
\item[] Apêndice D: Equilíbrio químico e a inclusão de elétrons na equação de estado.
\end{itemize}

\begin{widetext}
  \newpage
  \section*{Apêndices}
  \subsection*{Apêndice A - Sistema de unidades naturais} 
\vspace{5mm}

Para trabalhar com física de partículas, as unidades presentes no Sistema Internacional (SI) não são as mais convenientes, pois os valores cotidianos são elevados demais nessas escalas. Desta forma, é usual utilizar-se o chamado sistema de unidades naturais, definidos a partir de $\hbar = c = k_B = \num{1}$, com $\hbar = h/2\pi$ onde $h$ é a constante de Planck, $c$ é a velocidade da luz no vácuo e $k_B$ é a constante de Boltzmann. Assim, todas as propriedades de interesse, como a massa das partículas e a temperatura, são expressas em termos de energia. A unidade de energia mais conveniente é o elétron-volt e os seus respectivos múltiplos. Normalmente, utiliza-se o \si{GeV} ($=\SI{e9}{eV}$). Os fatores de conversão entre as unidades escritas no SI para o sistema natural de unidades estão apresentados na Tab.~\ref{tab:a}. Uma consequência direta desse sistema de unidades é que a relação de de Broglie, dada por $p = \hbar k$ , onde $p$ é o \textit{momentum} e $k$ é o número de onda, se torna $p = k$. Neste trabalho, $k$ possui unidades de \textit{momentum}, que em unidades naturais é \si{GeV}, e será referido simplesmente como \textit{momentum}. Todas as equações apresentadas no texto estão em unidades naturais. Nesse sistema de unidades, temos que pressão e densidade de energia são medidos em \si{GeV}$^4$, enquanto no SI são dados em \si{J\cubic\meter} . Como a conversão de fentômetro (\si{fm} $=$ \SI{e-15}{m}) para \si{GeV} é direta e dada por \SI{1}{fm} $=$ \SI{5,07}{GeV}$^{-1}$, faremos uso alternado, porém explícito, do sistema natural para a unidade de pressão e densidade de energia escrita como \si{GeV/fm}$^3$. Apresentaremos a densidade bariônica em termos de \si{fm}$^{-3}$ . Além disso, nos resultados apresentaremos o raio da estrela em \si{km} e sua massa em unidades de massa solar (M$_\odot$).

\begin{table*}[!h]
  \centering
  \caption{Fatores de conversão das unidades do SI para as unidades naturais.}
  \label{tab:a}
  \begin{tabular}{c c c}  
    \hline
    Fator de conversão & Unidades naturais ($\hbar = c = k_B = \num{1}$) & Dimensão verdadeira \\ \hline
    \SI{1}{kg} $=$ \SI{5,61e26}{GeV} & \si{GeV} & \si{GeV/\squared c} \\ \hline
    \SI{1}{m} $=$ \SI{5,07e15}{GeV}$^{-1}$ & \si{GeV^{-1}} & \si{\planckbar c/GeV} \\ \hline
    \SI{1}{s} $=$ \SI{1,52e24}{GeV}$^{-1}$ & \si{GeV^{-1}} & \si{\planckbar c/GeV} \\ \hline
    \SI{1}{K} $=$ \SI{8,62e-14}{GeV} & \si{GeV} & \si{GeV/}$k_B$\\ \hline    
  \end{tabular}
\end{table*}

\newpage

\subsection*{Apêndice B -- Equação de Tolman-Oppenheimer-Volkoff}
A equação TOV é derivada a partir das equações de Einstein para o campo gravitacional~\cite{oppenheimer1939,tolman1939} e dadas por (em unidades geométricas, onde $c = G = \num{1}$ e $G$ é a constante gravitacional)
\begin{equation}
  \label{eq:einstein}
  G_{\mu\nu} = -8\pi T_{\mu\nu}\,.
\end{equation}

O tensor métrico, presente no tensor de Einstein $G_{\mu\nu}$, pode ser obtido a partir do elemento de linha, que para uma estrela esfericamente simétrica e estática é dado por
\begin{equation}
  \label{eq:lineelement}
  \mathrm{d}s^2=e^{2\nu(r)}\mathrm{d}t^2-e^{2\lambda(r)}\mathrm{d}r^2-r^2[\mathrm{d}\theta^2+\sin^2\theta\, \mathrm{d}\phi^2]\;,
\end{equation}
tal que, com $\nu \equiv \nu (r)$ e $\lambda \equiv \lambda (r)$ temos
$$
\begin{matrix}
  g_{00}=e^{2\nu}\,; 
  &\,g_{11}=-e^{2\lambda}\,; \\
  \,g_{22}=-r^2\,;
  &\,g_{33}=-r^2\sin^2 \theta \;.
\end{matrix}
$$
Os símbolos de Christoffel são definidos a partir da métrica pela expressão a seguir
\begin{equation}
  \label{eq:christoffelgeral}
  \Gamma_{\mu\nu}^{\gamma}=\frac{1}{2}g^{\gamma\alpha}\left(\partial_{\mu} g_{\alpha\nu} + \partial_\nu g_{\mu\alpha} - \partial_\alpha g_{\mu\nu}\right)\;.
\end{equation}
Como o tensor métrico é diagonalizado, só temos termos não nulos quando $\gamma = \alpha$, assim
\begin{equation}
  \label{eq:christoffeldiag}
  \Gamma_{\mu\nu}^{\gamma}=\frac{1}{2}g^{\gamma\gamma}\left(\partial_{\mu} g_{\gamma\nu} + \partial_\nu g_{\mu\gamma} - \partial_\gamma g_{\mu\nu}\right)\;,
\end{equation}
o que resulta nos seguintes termos não nulos
$$
\begin{matrix}
  \Gamma^0_{10}= \Gamma^0_{01} = \nu '\;,
  &\Gamma^1_{00}= \nu 'e^{2(\nu - \lambda)}\;, 
  &\Gamma^1_{11}=\lambda ' \;,\\
  \Gamma^1_{22}=-r e^{-2\lambda} \;,
  &\Gamma^1_{33}=-r\sin^2\theta e^{-2\lambda} \;,
  &\Gamma^2_{12}=\Gamma^2_{21}=\dfrac{1}{r} \;,\\
  \Gamma^2_{33}=-\sin \theta \cos \theta \;,
  &\Gamma^3_{13}=\Gamma^3_{31} = \dfrac{1}{r} \;,
  &\Gamma^3_{23}=\Gamma^3_{32} = \cot \theta \;,\\
\end{matrix}
$$
onde a linha denota a derivada parcial com relação à $r$.

O tensor da curvatura de Riemann, na sua forma covariante, pode ser obtido a partir dos símbolos de Christoffel, sendo dado por
\begin{align}
  \label{eq:covriemann}
  R_{\delta\gamma\mu\nu} = {} & g_{\delta\mu}R^\alpha_{\gamma\mu\nu} \nonumber\\
  = {} & \frac{1}{2}(\partial_\gamma\partial_\mu g_{\delta\nu}+\partial_\delta\partial_\nu g_{\delta\mu} - \partial_\delta \partial_\mu g_{\gamma\nu}) + g_{\beta\eta}(\Gamma^\beta_{\gamma\mu}\Gamma^\eta_{\delta\nu}-\Gamma^\beta_{\gamma\nu}\Gamma^\eta_{\delta\mu})\;.
\end{align}
Portanto, o tensor de Ricci, que é definido por
\begin{equation}
  \label{eq:riccitensor}
  R_{\gamma\mu}=g^{\delta\nu}R_{\delta\mu\gamma\nu}\;,
\end{equation}
possui as seguintes componentes diagonais (as restantes são nulas por causa do tensor métrico): 
\begin{equation}
  \bullet ~ R_{00}=g^{00}R_{0000}+g^{11}R_{1001}+g^{22}R_{2002}+g^{33}R_{3003}\;,
\end{equation}
onde
\begin{align}
  R_{0000} = {} & 0\;,\\
  R_{1001} = {} & R_{0110}=(\nu '' + \nu '^2-\nu ' \lambda ')e^\nu \;, \\
  R_{2002} = {} & R_{0220}=r\nu 'e^{2(\nu-\lambda)}\;, \\
  R_{3003} = {} & R_{0330}=r\sin^2\theta\, \nu' e^{2(\nu-\lambda)}\;,
\end{align}
logo,
\begin{equation}
  \label{eq:r00}
  R_{00}=\left(\nu''-\nu'^2+\nu'\lambda' - 2\frac{\nu'}{r}\right)e^{2(\nu-\lambda)}\,.
\end{equation}

\begin{equation}
  \bullet ~ R_{11}=g^{00}R_{0110}+g^{11}R_{1111}+g^{22}R_{2112}+g^{33}R_{3113}\,,
\end{equation}
com $R_{0110}$ já calculado, temos
\begin{align}
  R_{1111}  = {} &  0\;, \\
  R_{2112} = {} &  R_{1221} = r\lambda'\;, \\
  R_{3113} = {} &  R_{1331} = r\sin^2 \theta \;\lambda' \;,
\end{align}
logo,
\begin{equation}
  \label{eq:r11}
  R_{11}= \nu'' + \nu'^2-\nu'\lambda'-2\frac{\lambda'}{r}\,.
\end{equation}

\begin{equation}
  \bullet ~ R_{22}=g^{00}R_{0220}+g^{11}R_{1221}+g^{22}R_{2222}+g^{33}R_{3223}\;,
\end{equation}
com $R_{0220}$ e $R_{1221}$ já calculados e $R_{2222}=0$, temos
\begin{equation}
  R_{3223}= R_{2332} = r^2\sin^2\theta(1-e^{-2\lambda})\;,
\end{equation}
assim
\begin{equation}
  \label{eq:r22}
  R_{22}=(r\nu'-r\lambda'+1)e^{-2\lambda}-1\,.
\end{equation}

\begin{equation}
  \bullet ~ R_{33}=g^{00}R_{0330}+g^{11}R_{1331}+g^{22}R_{2332}+g^{33}R_{3333}\,,
\end{equation}
como $R_{3333}=0$ e com os outros coeficientes já calculados anteriormente, chegamos a
\begin{equation}
  \label{eq:r33}
  R_{33}=R_{22}\sin^2\theta=[(r\nu'-r\lambda'+1)e^{-2\lambda}-1]\sin^2\theta\,.  
\end{equation}

A partir do tensor de Ricci, somos capazes de determinar o escalar de Ricci, tal que
\begin{equation}
  \label{eq:escricci1}
  R = g^{\mu\nu}R_{\mu\nu}\;,
\end{equation}
o que fornece
\begin{equation}
  \label{eq:escricci}
  R = 2\left(-\nu''-\nu'^2+\nu'\lambda'-2\frac{\nu'-\lambda'}{r}-\frac{1}{r^2}\right)e^{-2\lambda}+\frac{2}{r^2}\;.
\end{equation}
Assim, podemos obter o lado esquerdo das equações de Einstein, isto é, o tensor de Einstein, que na sua forma mista é dado por
\begin{equation}
  G^\mu_\nu=R^\mu_\nu-\frac{1}{2}\delta^\mu_\nu R\;,
\end{equation}
onde $\delta^\mu_\nu$ é a delta de Kronecker, que possui valor igual a um quando $\mu = \nu$ e zero para $\mu \neq \nu$. Portanto, temos
\begin{align}
  G^0_0 = {} & \left(\frac{1}{r^2}-2\frac{\lambda'}{r}\right)e^{-2\lambda}-\frac{1}{r^2}\;, \\
  G^1_1 = {} & \left(\frac{1}{r^2}+2\frac{\nu'}{r}\right)e^{-2\lambda}-\frac{1}{r^2}\;, \\
  G^2_2 = {} &  G^3_3=\left(\nu''+\nu'^2 - \nu'\lambda'+\frac{\nu'-\lambda'}{r}\right)e^{-2\lambda}\;.  
\end{align}

O lado direito da equação de Einstein é obtido ao considerarmos que a estrela é composta por um fluido ideal e isotrópico, sendo assim, o tensor energia-\textit{momentum} na sua forma mista é escrito
\begin{equation}
  T^\mu_\nu=(\epsilon + p)u^\mu u_\nu - p\delta^\mu_\nu \;,
\end{equation}
onde $u^\mu$ é o quadrivetor velocidade ou quadrivelocidade do fluido. Desta forma, no referencial de repouso do fluido $u^\mu=(1,0,0,0)$. Por isso, temos
\begin{equation}
  T^0_0=\epsilon\;,\;T^1_1=T^2_2=T^3_3=-p\;.
\end{equation}

A partir da aplicação dos tensores $G^\mu_\nu$ e $T^\mu_\nu$ na Eq.~(\ref{eq:einstein}), obtemos as seguintes relações
\begin{align}
  \label{eq:G00}
  \left(\frac{1}{r^2}-\frac{2\lambda'}{r}\right)e^{-2\lambda}-\frac{1}{r^2}  = {} &  - 8\pi\epsilon(r)\,, \\
  \label{eq:G11}
  \left(\frac{1}{r^2}+\frac{2\nu'}{r}\right)e^{-2\lambda} - \frac{1}{r^2}  = {} &  8\pi p(r)\,, \\
  \label{eq:G22}
  \left(\nu'' + \nu'^2-\lambda'\nu' + \frac{\nu' - \lambda'}{r}\right)e^{-2\lambda}  = {} & 8\pi p(r)\,.
\end{align}

Podemos relacionar $e^{2\lambda}$ com a massa da estrela já que
\begin{equation}
  \label{eq:ddr}
  \frac{\mathrm{d}}{\mathrm{d}r}\left[r\left(1-e^{-2\lambda}\right)\right]= 1-e^{-2\lambda}+2r\,\lambda'e^{-2\lambda}\,,
\end{equation}
o que, multiplicando os dois lados da equação por $r^{-2}$, resulta em
\begin{equation}
  \label{eq:1r2}
  \frac{1}{r^2}\frac{\mathrm{d}}{\mathrm{d}r}\left[r\left(1-e^{-2\lambda}\right)\right] = - \left[e^{-2\lambda}\left(\frac{1}{r^2}-\frac{2\lambda'}{r}\right) - \frac{1}{r^2}\right]\,.
\end{equation}
Assim, a partir da Eq.~(\ref{eq:G00}), temos
\begin{equation}
  8\pi\epsilon(r)=\frac{1}{r^2}\frac{\mathrm{d}}{\mathrm{d}r}\left[r\left(1-e^{-2\lambda}\right)\right]\,,
\end{equation}
ou, multiplicando por $r^2$ e integrando, temos
\begin{equation}
  8\pi\int^r_0r'^2\epsilon(r')\mathrm{d}r' = r(1-e^{-2\lambda}) \,.
\end{equation}
Isolando $e^{-2\lambda}$ temos
\begin{equation}
  \label{eq:e-2lambda}
  e^{-2\lambda} = 1 - \frac{8\pi}{r}\int^r_0r'^2\epsilon(r')\mathrm{d}r'\,.
\end{equation}

Como a relação entre massa e raio no interior da estrela é descrita por
\begin{equation}
  \label{eq:dmdr2}
  \frac{\mathrm{d}m}{\mathrm{d}r} = 4\pi r^2\epsilon(r)\,,
\end{equation}
que na sua forma integral fica
\begin{equation}
  \label{eq:mr}
  m(r) = 4\pi\int^r_0r'^2\epsilon(r')\mathrm{d}r'\,,
\end{equation}
podendo ser associada com a Eq.~(\ref{eq:e2lambda}), tal que
\begin{equation}
  \label{eq:e2lambda}
  e^{2\lambda} = \left(1-\frac{2 m(r)}{r}\right)^{-1}\,.
\end{equation}

Ainda pela Eq.~(\ref{eq:G00}), podemos isolar $\lambda'$, o que resulta em
\begin{equation}
  \label{eq:lambda'}
  \lambda' = \frac{1}{2r}\left\{1 - e^{2\lambda}\left[1-8\pi r^2\epsilon(r)\right]\right\}\,.
\end{equation}

Na Eq.~(\ref{eq:G11}), podemos isolar $\nu'$, obtendo
\begin{equation}
  \label{eq:nu'}
  \nu' = \frac{1}{2r}\left\{e^{2\lambda}\left[8\pi r^2p(r)+1\right]-1\right\}\,.
\end{equation}

A fim de resolvermos Eq.~(\ref{eq:G22}), devemos calcular, a partir da Eq.~(\ref{eq:nu'}), $\nu'^2$ e $\nu''$, além do produto $\nu'\lambda'$ e da diferença ($\nu' - \lambda'$)/$r$.
Após cálculos diretos, porém longos, chegamos aos seguintes resultados:
\begin{align}
  \label{eq:nu'2}
  \nu'^2 = {} & e^{4\lambda}\left[16\pi^2r^2p^2(r)+4\pi p(r) + \frac{1}{4r^2}\right] - e^{2\lambda}\left[4\pi p(r) + \frac{1}{2r^2}\right] + \frac{1}{4 r^2}\,,\\
  \label{eq:nu''}
  \nu'' = {} & e^{4\lambda}\left[32\pi^2r^2\epsilon(r)p(r)+4\pi\epsilon(r) -4\pi p(r)- \frac{1}{2r^2}\right] + e^{2\lambda}[4\pi r p'(r)+8\pi p(r)] + \frac{1}{2r^2} \,,  \\
  \label{eq:nu'lambda'}
  \nu'\lambda' = {} & e^{4\lambda}\left[16\pi^2r^2p(r)\epsilon(r)+2\pi\epsilon(r)-2\pi p(r) - \frac{1}{4 r^2}\right] + e^{2\lambda}\left[2\pi p(r) - 2\pi\epsilon(r) + \frac{1}{2 r^2}\right] - \frac{1}{4r^2}\,, \\
  \label{eq:nu'-lambda'}
  \frac{\nu' - \lambda'}{r} = {} & e^{2\lambda}\left[4\pi p(r)-4\pi\epsilon(r)+\frac{1}{r^2}\right] - \frac{1}{r^2}\,.
\end{align}
Substituindo os resultados acima na Eq.~(\ref{eq:G22}), agrupando os termos e simplificando-os, temos
\begin{equation}
  \{e^{2\lambda}[8\pi r^2p(r)+1] - 1\}[\epsilon(r) + p(r)] + 2r\,p'(r) = 0\,.
\end{equation}
Isolando $p'(r) = \mathrm{d}p/\mathrm{d}r$ e utilizando o resultado obtido na Eq.~(\ref{eq:e2lambda}), finalmente, chegamos à equação TOV em unidades geométricas, dada por
\begin{equation}
  \label{eq:tovvv}
  \frac{\mathrm{d}p}{\mathrm{d}r}=-\frac{m(r)\epsilon(r)}{r^2}\left[1+\frac{p(r)}{\epsilon(r)}\right]\left[1+\frac{4\pi r^3 p(r)}{m(r)}\right]\left[1-\frac{2m(r)}{r}\right]^{-1}\,.
\end{equation}

\newpage

\subsection*{Apêndice C - Temperatura e potencial bariônico críticos na transição de fase de primeira ordem}
{Neste apêndice iremos derivar os valores críticos da temperatura e do potencial químico bariônico considerando o modelo de sacola do MIT e que a transição de fase seja de primeira ordem. Como enfatizado anteriormente, ambos pressupostos são aproximações rudimentares da realidade. Entretanto, nos possibilitam obter uma estimativa da ordem de grandeza em que se espera a transição de fase entre a matéria de hádrons e àquela de quarks} 
\vspace{5mm}

Seja um plasma de quarks e glúons, contendo apenas os sabores de quarks \textit{up} e \textit{down}. Temos que a pressão do gás, com potencial químico nulo, será descrita por [13]
\begin{equation}
  \label{eq:pdpa}
  P = \frac{37\pi^2}{90}T^4\,.
\end{equation}
Neste caso, a temperatura crítica para que ocorra a transição de fase de primeira ordem é obtida quando a pressão do plasma for igual à pressão de sacola $B$. Desta forma, temos que
\begin{equation}
  \label{eq:tccc}
  T_c = \left(\frac{90}{37\pi^2}\right)^{1/4}B^{1/4} \,.
\end{equation}
Para $B^{1/4} = \SI{206}{MeV}$ temos que $T_c \sim \SI{144}{MeV}$. Se o plasma de quarks e glúons for submetido a temperaturas maiores, ocorrerá o desconfinamento, caracterizando uma matéria de quarks livres.

A relação termodinâmica entre a pressão e o potencial químico do gás de quarks (contendo apenas dois sabores), a temperatura zero, é tal que
\begin{equation}
  \label{eq:pqpq}
  P_q = \frac{1}{2\pi^2}\mu_q^4\,.
\end{equation}
Novamente, a mudança de estado na matéria ocorrerá quando $P_q = B$, o que nos leva a
\begin{equation}
  \mu_q = \left(2\pi^2B\right)^{1/4}\,.
\end{equation}

A relação entre a densidade bariônica crítica e o potencial químico é tal que
\begin{align}
  n_c = &\; {} \frac{2}{3\pi^2}\mu_q^3 = \frac{2}{3\pi^2}\left(2\pi^2 B\right)^{3/4} \nonumber \\
  = &\; {} 4\left(\frac{1}{2\pi^2}\right)^{1/4}B^{3/4}\,.
\end{align}
Considerando $B^{1/4} = \SI{206}{MeV}$ obtemos uma densidade bariônica crítica de $n_c = \SI{0,72}{fm^{-3}}$. Sendo a densidade bariônica nuclear $n_0 = \SI{0,16}{fm^{-3}}$, temos que $n_c = \num{4,5}\,n_0$.

\newpage
\subsection*{Apêndice D - Equilíbrio químico e a inclusão de elétrons na equação de estado}
Neste apêndice, iremos demonstrar como a inclusão de elétrons na equação de estado é necessária para que tenhamos a matéria estranha de quarks em equilíbrio químico (também chamado de equilíbrio-$\beta$) e neutralidade elétrica. No texto principal, ao considerarmos quarks massivos, simplificamos o problema por assumir que o \textit{momentum} de Fermi dos três quarks era igual e desconsideramos a presença de elétrons na estrela, o que não satisfaz o equilíbrio-$\beta$ e a neutralidade elétrica simultaneamente. No caso do equilíbrio-$\beta$, temos que as reações fracas implicam em
\begin{equation}
  \label{eq:quimico}
  \mu_d = \mu_u + \mu_{e^-} \quad\mathrm{e}\quad \mu_d = \mu_s\,.
\end{equation}

Por outro lado, a condição da neutralidade elétrica implica
\begin{equation}
  \label{eq:neutralidade}
  q = \sum_f Q_f n_f = \frac{2}{3}n_u - \frac{1}{3}(n_d + n_s) - n_{e^-} = 0\,,
\end{equation}
onde $n_f = g_fk_f^3/(6\pi^2)$ e $g_f$ é o fator de degenerescência da partícula $f(=u,d,s,e^-)$ sendo igual a dois para os léptons e igual a seis para os quarks. Para partículas massivas temos que $k_f = (\mu_f - m_f)^{1/2}$, isto implica que a Eq.~(\ref{eq:neutralidade}) pode ser reescrita como
\begin{equation}
  \label{eq:neutralidade2}
  2(\mu_u^2 - m_u^2)^{3/2} - [(\mu_d^2 - m_d^2)^{3/2} + (\mu_s^2 - m_s^2)^{3/2} + (\mu_{e^-}^2 - m_{e^-}^2)^{3/2}] = 0\,.
\end{equation}

Dos quatro potenciais químicos relacionados na Eq.~(\ref{eq:quimico}), temos que apenas dois são variáveis independentes. Sendo assim, a condição de neutralidade elétrica global nos fornece a última condição necessária para relacioná-los e resolvermos o sistema de equações. Numericamente, consideramos que o potencial químico do quark strange (que é igual ao do quark down) é a variável independente. Utilizando um método de encontrar raízes, estabelecemos pela condição de neutralidade elétrica global o potencial químico dos elétrons.

Tais considerações levam a uma mudança na equação de estado e na forma numérica de resolvê-la. A pressão e a densidade de energia do gás passam a ser dadas, respectivamente, por
\begin{align}
  \label{eq:MITp}
  p = {} & - B + \sum_f\frac{g_f}{24\pi^2}\left[\mu_fk_f\left(\mu_f^2-\frac{5}{2}m_f^2\right) + \frac{3}{2}m_f^4\ln\left(\frac{\mu_f+k_f}{m_f}\right)\right] \,,  \\
  \label{eq:MITe}
  \epsilon = {} & B +\sum_f\frac{g_f}{8\pi^2}\left[\mu_fk_f\left(\mu_f^2-\frac{1}{2}m_f^2\right) - \frac{1}{2}m_f^4\ln\left(\frac{\mu_f+k_f}{m_f}\right)\right]\,,
\end{align}
que são iguais as expressões do texto principal com a adição do fator de degenerescência e do termo para os elétrons. Este termo adicional para os elétrons suaviza a equação de estado, gerando estrelas com massas e raios menores do que na nossa aproximação, onde só quarks populam a estrela, como pode ser visto na Fig.~\ref{MxR-e}.

\begin{figure*}[!b]
  \centering
  \includegraphics[width=.4\textwidth]{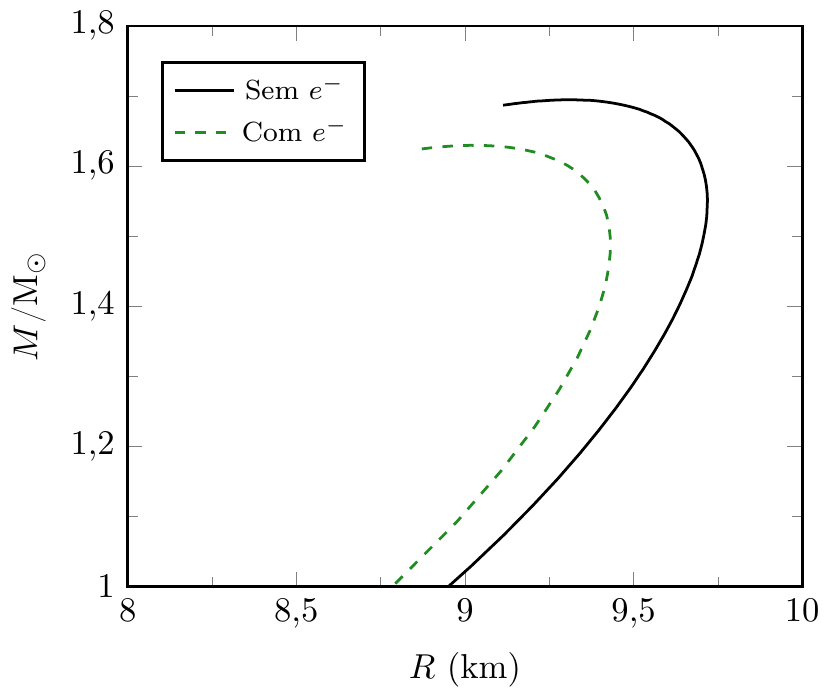}
  \caption{\label{MxR-e} Perfil massa-raio para estrelas estranhas considerando e desconsiderando a presença de elétrons.}
\end{figure*}

\end{widetext}

\end{document}